\documentclass[12pt,a4paper]{article}
\usepackage{afterpage}
\usepackage{epsfig}
\usepackage[a4paper]{geometry}
\usepackage{float}
\usepackage[stable]{footmisc}
\usepackage[utf8x]{inputenc}
\usepackage{a4wide}
\usepackage{amsmath,amssymb,amstext,amsthm}
\usepackage{amsfonts}
\usepackage{framed}
\usepackage{graphicx}
\usepackage{subcaption}
\usepackage{setspace}
\usepackage{bbold}
\usepackage{bbm}
\usepackage{array}
\usepackage{hyperref}
\usepackage{overpic}
\usepackage{cite}
\usepackage{lipsum}
\usepackage{color}
\usepackage{enumitem}
\usepackage{cancel}
\usepackage{verbatim}
\usepackage{amsthm} 
\usepackage{hyperref}
\newtheorem{Theorem}{Theorem}

\graphicspath{{figures/}}

\newcommand{\del}[0]{\partial}
\let\baraccent=\=
\renewcommand{\=}[1]{\stackrel{#1}{=}}
\newcommand{\id}[0]{\mathbb{1}}
\begin{document}
	\pagestyle{plain}

	%----------------------------------------------------------------------%
	%  numbering equations with section number
	%----------------------------------------------------------------------%
	\makeatletter
	\@addtoreset{equation}{section}
	\makeatother
	\renewcommand{\theequation}{\thesection.\arabic{equation}}
	%----------------------------------------------------------------------%
	%  title page
	%----------------------------------------------------------------------%
	\pagestyle{empty}
	%\vspace*{1.0in}
	\rightline{DESY 20-046}
	%\rightline{\tt hep-th/yymmnnn}
	\vspace{0.5cm}
	\begin{center}
		\Huge{{A landscape of orientifold vacua}
			\\[15mm]}
		\normalsize{Federico Carta,$^{\dagger }\footnote{\text{federico.carta@desy.de}}$ Jakob Moritz,$^{\dagger\dagger}\footnote{\text{moritz@cornell.edu}}$ and Alexander Westphal$^{\dagger}\footnote{\text{alexander.westphal@desy.de}}$ \\[10mm]}
		\small{\slshape
			$^{\dagger}$ Deutsches Elektronen-Synchrotron, DESY, Notkestra$\beta$e 85,\\ 22607 Hamburg, Germany. \\[2mm]  } 
		\small{\slshape
			$^{\dagger \dagger}$ Department of Physics, Cornell University, Ithaca, NY 14853, USA \\[4mm]  }
		
		\normalsize{\bf Abstract} \\[4mm]
	\end{center}
	\begin{center}
		\begin{minipage}[h]{15.0cm} 
			We present a vast landscape of O3/O7 orientifolds that descends from the famous set of complete intersection Calabi-Yau threefolds (CICY). We give distributions of topological data relevant for phenomenology such as the orientifold-odd Hodge numbers, the D3-tadpole, and multiplicities of O3 and O7-planes. Somewhat surprisingly, almost all of these orientifolds have conifold singularities whose deformation branches are projected out by the orientifolding. However, they can be resolved, so most of the orientifolds actually descend from a much larger and possibly new set of CY threefolds that can be reached from the CICYs via conifold transitions. We observe an interesting class of $\mathcal{N}=1$ geometric transitions involving colliding O-planes. Finally, as an application, we use our dataset to produce examples of orientifolds that satisfy the topological requirements for the existence of ultra-light throat axions (\textit{thraxions}) as proposed in \cite{Hebecker:2018yxs}. The database can be accessed \href{https://www.desy.de/~westphal/orientifold_webpage/cicy_orientifolds.html}{here}.

		\end{minipage}
	\end{center}
	\newpage
	%----------------------------------------------------------------------%
	%  Resetting of counters
	%----------------------------------------------------------------------%
	\setcounter{page}{1}
	\pagestyle{plain}
	\renewcommand{\thefootnote}{\arabic{footnote}}
	\setcounter{footnote}{0}
	%----------------------------------------------------------------------%
	%  Paper begins
	%----------------------------------------------------------------------%
	
\tableofcontents
\newpage

\section{Introduction and Conclusions}

The landscape of flux vacua in the type IIB corner of critical string theory \cite{Giddings:2001yu} is a fruitful arena for model building and for addressing fundamental issues such as moduli stabilization, and the existence of de Sitter vacua in string theory. The landscape is believed to be vast \cite{Susskind:2003kw,Ashok:2003gk,Denef:2004ze}, giving rise to a plethora of low energy effective field theories (EFT). While their spectrum and Wilson coefficients are expected to vary seemingly randomly along the bulk of the landscape it has been conjectured that the landscape populates only small islands within a (much) larger space of EFTs that are not realized in string theory, known as the swampland \cite{Vafa:2005ui,Ooguri:2006in}. Clearly, by charting out the boundary of the landscape one hopes to understand how string theory constrains the set of low energy observables. A somewhat more modest and related goal is to understand how well distinct field theory sectors can be decoupled from one-another at the level of non-renormalizable operators, and how weak their (renormalizable) self-interactions can be tuned. Due to the Dine-Seiberg problem \cite{Dine:1985he} this is related to how far the landscape extends away from its bulk.\footnote{Examples of \textit{decoupling} parameters include the magnitude of the flux superpotential in KKLT \cite{Kachru:2003aw}, and backreaction radii of branes in comparison to the overall volume of the CY (see e.g.\cite{Freivogel:2008wm,Carta:2019rhx} in the same context).}

This motivates us to study on a grand scale the CY orientifolds of O3/O7 type which are a basic starting point for constructing phenomenologically interesting vacua: Consider a choice of Calabi-Yau (CY) threefold $X$ together with a holomorphic involution $\mathcal{I}: X\rightarrow X$ that acts as $(-1)$ on the holomorphic threeform $\Omega$ of $X$. This data determines a type IIB O3/O7 orientifold with O3 and O7 planes residing at the connected components of the geometric fixed point locus $\mathcal{F}$ of $\mathcal{I}$ in $X$ of co-dimension three and one respectively (see e.g. \cite{Polchinski:1998rr}). The Ramond-Ramond (RR) tadpoles induced by the O-planes can be canceled by introducing appropriate configurations of D7 branes with or without world-volume fluxes, bulk three-form fluxes, and/or mobile D3 branes. This generates $4d$ $\mathcal{N}=1$ low energy effective supergravity theories featuring perturbative no-scale vacua with spontaneously broken supersymmetry, non-abelian gauge theories, chiral sectors etc (see e.g. \cite{Ibanez:2012zz,Grana:2005jc}). An important subset of the data determining these effective field theories is given by simple topological data of $X$ and its involution $\mathcal{I}$. For example, the dimensions of the orientifold even/odd cohomology groups $H^{2,1}_{-},H^{1,1}_+,H^{2,1}_+,H^{1,1}_-$ determine the number of complex structure moduli, K\"ahler moduli, closed string $U(1)$ vector multiplets, and axionic chiral multiplets respectively. For simplistic choices of D7 brane configurations, the number of connected components of $\mathcal{F}$ of co-dimension one sets the number of non-abelian gauge sectors with gauge algebra so$(8)$, and the Euler characteristic of $\mathcal{F}$ restricts the freedom to choose world volume and bulk three-form fluxes.

In this article, we construct a database of 2,004,513 such orientifolds, and compute a number of topological invariants that can be used as input data for phenomenological model building. These include the D3-tadpole, the dimensions of the orientifold-odd cohomology groups, the number of O3 and O7 planes, and topological data of O7-divisors. The starting point is the classic database of complete intersection CY (CICY) threefolds, as constructed by Candelas et al \cite{Candelas:1987kf}. We make use of their most tractable\footnote{The descriptions are useful because the divisor classes of the CY threefold are simply inherited by the ones of the ambient space. Only about half of the descriptions in the original database \cite{Candelas:1987kf} of Candelas et al have this property.} descriptions, obtained by Anderson et al in \cite{Anderson:2017aux}, \textit{either} as a complete intersection of hypersurfaces in certain products of projective spaces (henceforth \textit{favourable} CICYs), \textit{or} del Pezzo surfaces or rational elliptic surfaces (henceforth \textit{non-favourable} CICYs). The database is constructed by finding all ambient space involutions, and all $\mathbb{Z}_2$-invariant deformation classes of CICY embeddings. For the non-favourable cases, we incorporate the involutions of del-Pezzo surfaces and rational elliptic surfaces found by Blumenhagen et al \cite{Blumenhagen:2008zz} and Donagi et al \cite{Donagi:2000fw} respectively. 

Most of the orientifolds we produce are singular at co-dimension three, with a number of conifold singularities residing on the O7 planes which cannot be deformed in a way that would be compatible with the orientifold projection. However, all of these have a number of distinct resolution branches that are orientifold preserving, so to each singular orientifold there belongs a number of different geometric phases related to each other via flop transitions. Across such transition loci, the number of O3 planes as well as the topology of O7 divisors jumps, in a way that preserves the D3 charge.\footnote{Similar transitions across which an O7 plane \textit{eats} O3 planes have previously been described in the literature \cite{Denef:2005mm}.} Thus, the number of distinct smooth geometric phases is actually much larger than the quoted $\sim 2\times 10^6$.

Since the resolution phases are not contained in the CICY database, we produce, as a byproduct, many CY threefolds that are connected to the CICY database via a number of conifold transitions, but are themselves not contained in the original CICY database. It would be interesting to understand how many of these are already contained in other lists of CYs such as the Kreuzer-Skarke database \cite{Kreuzer:2000xy}, and how many are new.

We present two applications of our database. First, in section \ref{sec::stat_results}, we display the distributions of topological data that we have obtained, comment on some of their statistical properties, display the boundary of our CICY-orientifold landscape and speculate about what features we expect to generalize beyond the CICY-orientifold landscape. Second, in section \ref{sec::throataxions}, we use our database to find many orientifolds that satisfy all topological requirements known to us to realize models of ultralight throat axions, as proposed in \cite{Hebecker:2015tzo,Hebecker:2018yxs}. We expect these to be an ideal framework for testing and challenging swampland conjectures such as the weak gravity conjecture \cite{ArkaniHamed:2006dz} (WGC) and the swampland distance conjecture \cite{Ooguri:2006in} (SDC).

This paper is organized as follows. In section \ref{sec::stat_results} we discuss the distributions of topological data that we have obtained and use this to speculate about the boundary between the landscape and the swampland. The bulk of our paper is section \ref{sec::construction} where we explain how the database is constructed. In section \ref{sec::throataxions} we show how to find examples with throat axions.

\section{Results}\label{sec::stat_results}
In this section we display interesting features of the distributions we have obtained. First, let us explain what topological properties we have computed for each orientifold, and why:

\begin{enumerate}
	\item The orientifold-odd Hodge number $h^{1,1}_-$: This determines the number of perturbatively massless axionic chiral multiplets $\mathcal{G}_i\equiv \int_{\Sigma_2^i}C_2-\tau B_2$ \cite{Grimm:2004uq}. Here, $C_2$ and $B_2$ are the two-form potentials of the ten-dimensional type IIB supergravity theory, and $\Sigma_2^i$ are the orientifold-odd two-cycles. These axions are promising inflaton-candidates in models of axion-monodromy \cite{Silverstein:2008sg} such as \cite{McAllister:2008hb,Hebecker:2018yxs}. A histogram is shown in figure~\ref{fig:hi1histo}.
	\begin{figure}%[t!]
		\centering
		\includegraphics[width=8cm]{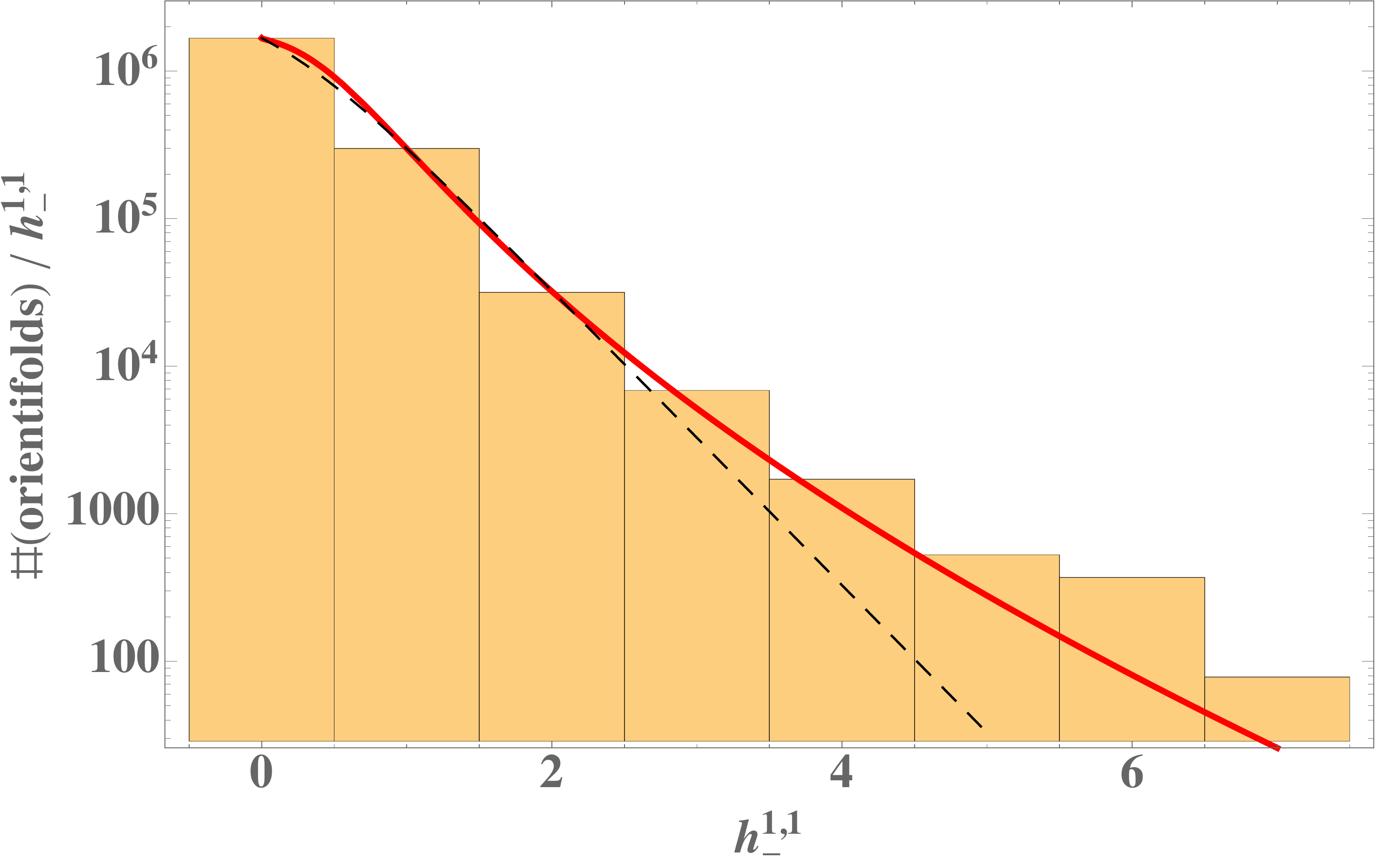}\includegraphics[width=8cm]{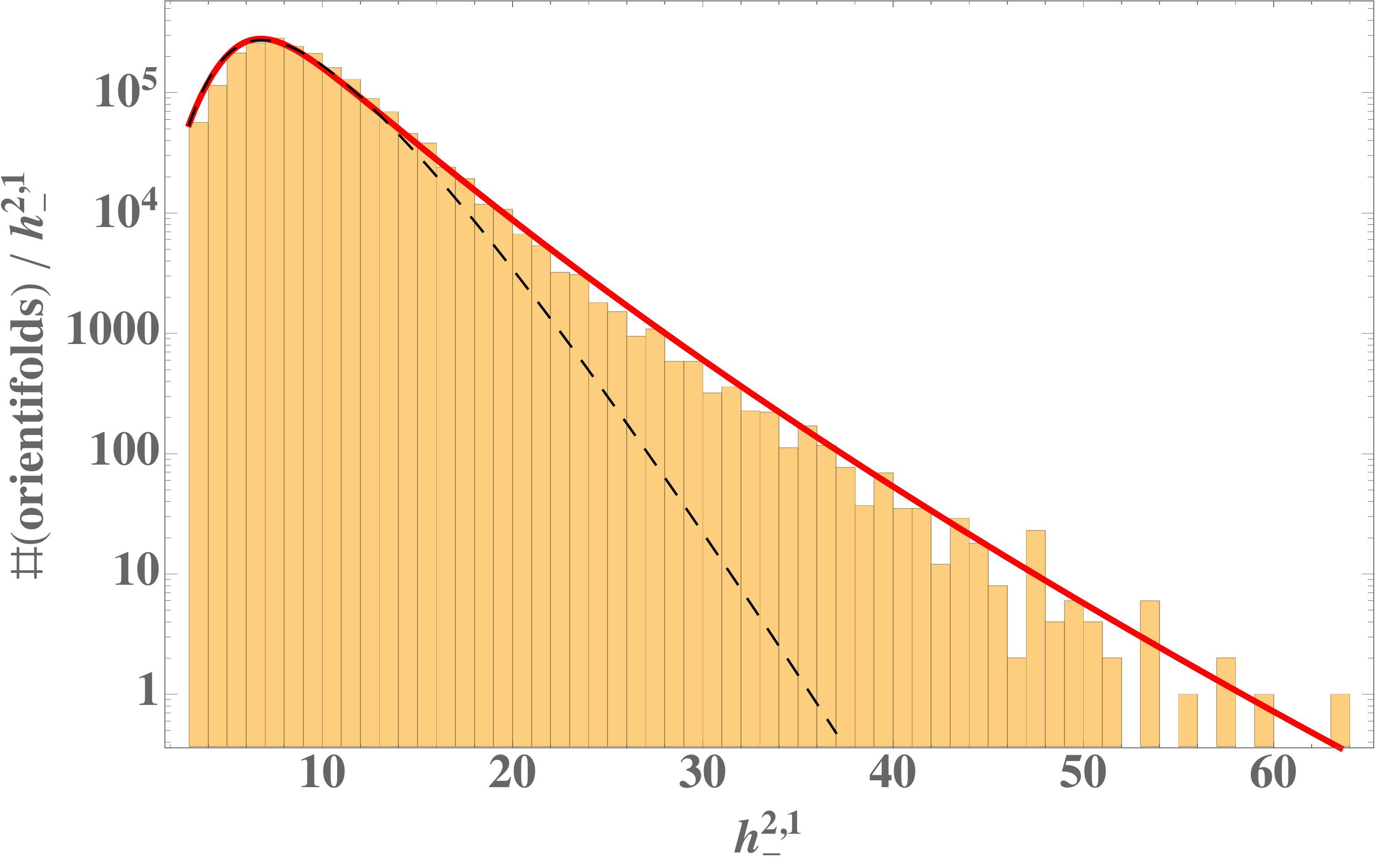}
		\caption{Histogram showing the number of orientifolds with given $h^{1,1}_-$ (left) and $h^{2,1}_-$ (right) on a logarithmic scale. Black dashed curve: Fit with $\#(h^{i,1}_-)\propto   (h^{i,1}_-)^{p_i}/(b_i+e^{c_i \,h^{i,1}_-})$, $p_1=0$. Red curve: Fit with $\#(h^{i,1}_-)\propto (h^{i,1}_-)^{p_i}/(b_i+e^{c_i \sqrt{h^{i,1}_-}})$, $p_1=0$. The latter fit results in parameter estimates $(b_1,c_1)\approx (71,5.8)$ and $(b_2,c_2,p_2)\approx (5.9\times 10^{-5},3.9,3.1)$. It is clear that the distributions of $h^{i,1}_-$ have a tail heavier than exponential fall-off (compare e.g. \cite{Long:2014fba}).}
		\label{fig:hi1histo}
	\end{figure}
	The fraction of orientifolds with $h^{1,1}_->0$ is about $17\%$.\footnote{See also \cite{Gao:2013pra,Altman} for results on orientifolds with $h^{1,1}_-\neq 0$ in the Kreuzer-Skarke database \cite{Kreuzer:2000xy}.}
	\item The orientifold-odd Hodge number $h^{2,1}_-$: This computes the number of bulk complex structure moduli that remain after the orientifold projection. These have to be stabilized by three-form fluxes \cite{Giddings:2001yu}. See figure~\ref{fig:hi1histo} for a histogram.
	\item The Euler characteristic of the CY threefold $\chi_{CY}=2(h^{1,1}-h^{2,1})$. For orientifolds of \textit{smooth} CICYs, the Euler characteristic \cite{Candelas:1987kf} and even the tuple $(h^{1,1},h^{2,1})$ of the underlying CY threefold is well-known \cite{Green:1987cr}. However, as mentioned in the introduction, most of the orientifolds that we determine are involutions of \textit{singular} CICYs with singularities that can be resolved in ways compatible with the orientifold projection. The smooth threefold thus obtained has an Euler characteristic different from the one of the CICY. Given (a) and (b) this also determines $h^{1,1}_+-h^{2,1}_+$, while the computation of the tuple $(h^{1,1}_+,h^{2,1}_+)$ is left for future work. The distribution of $\chi_{CY}$ is shown in figure~\ref{fig:chihisto}.
	\begin{figure}%[t!]
\centering
\includegraphics[width=10cm]{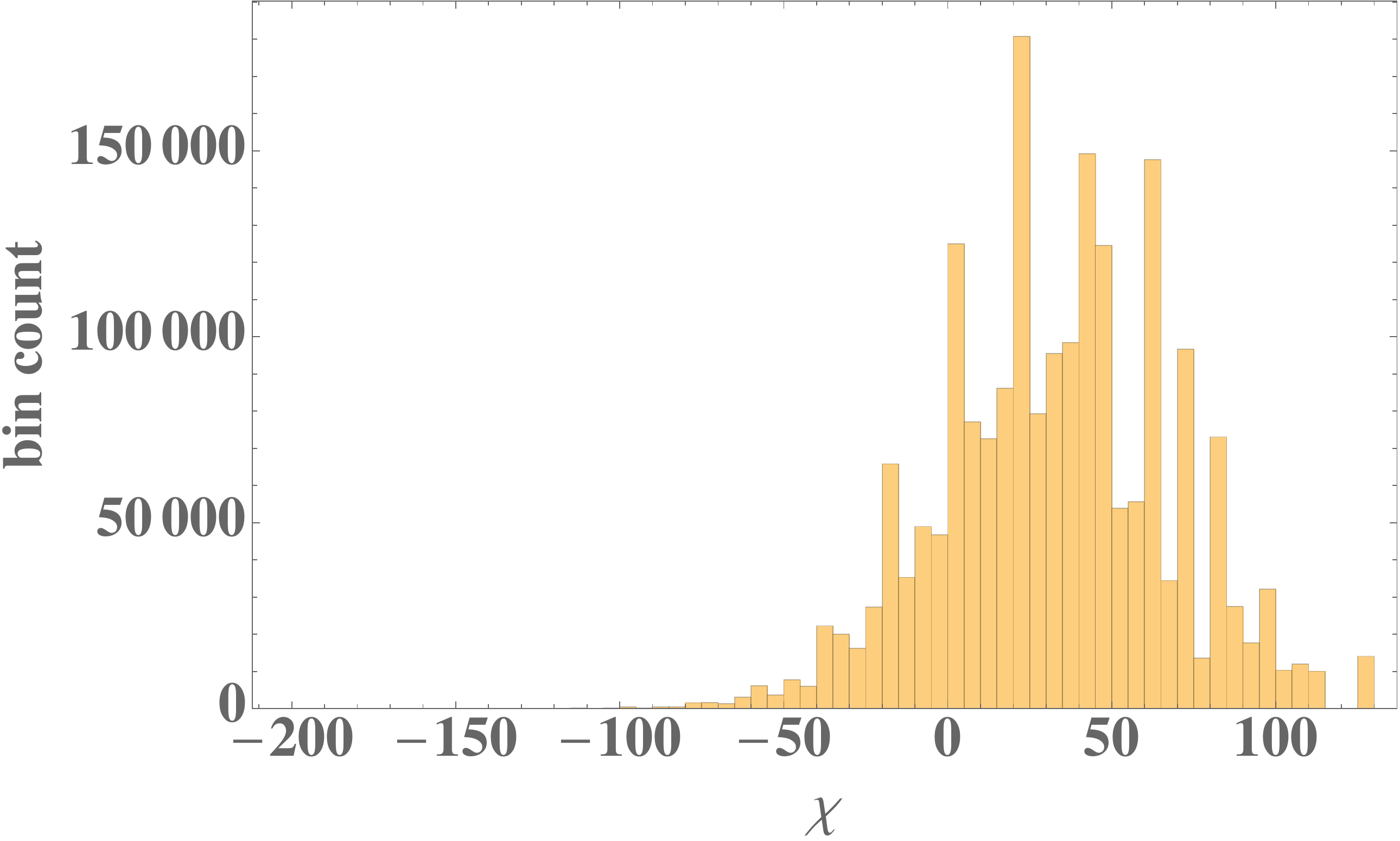}
\caption{Histogram showing the number of orientifolds with given Euler characteristic $\chi$ (of the CY threefold) with bin width $5$. The most negative value occurs for the quintic with $\chi=-200$. Given that $\chi$'s of both signs appear it is tempting to speculate that our list of CYs contains pairs related by mirror symmetry. The mean is shifted to positive values, because we reach smooth CYs by going through conifold transitions from the deformed to the resolved side, strictly increasing $\chi$.}
\label{fig:chihisto}
\end{figure}
	
	\item The set of distinct O-planes: The O7 planes contribute a D7-tadpole that can be canceled by introducing D7-branes. Stacks of seven-branes host non-abelian gauge groups, with charged matter living on intersection curves, and chiral index set by world volume fluxes, relevant for particle Physics model building (see e.g. \cite{Weigand:2018rez}). Furthermore, $\mathcal{N}=1$ pure Yang-Mills (YM) sectors confine and generate scalar potentials for K\"ahler moduli. For each O7 plane, we compute the Euler characteristic of the wrapped divisor $\chi_D$, and its \textit{degree} $d_D\equiv \int_D c_1(D)^2$. These determine the induced D3 charge dissolved in the 7-branes, and the \textit{arithmetic genus} $\chi(D,\mathcal{O}_D)$. Divisors with $\chi(D,\mathcal{O}_D)=1$ may generate non-perturbative superpotential terms, via euclidean D3 brane instantons, or the above mentioned strong gauge dynamics \cite{Witten:1996bn}. In figure~\ref{fig:nO7nO3histos} we display the distributions of the number O7 planes and the \textit{minimal} and \textit{maximal} number of O3 planes.\footnote{As explained in section \ref{sec:conifold}, for each singular CICY there are many resolution branches which differ from each other by the number of extra O3 planes that reside on the exceptional curves. The minimal number of O3 planes corresponds to a resolution branch that produces no extra O3 planes, while the maximal number of O3 planes corresponds to a resolution branch with the maximal number of extra O3 planes equal to the number of resolved conifolds.}
	\begin{figure}%[t!]
		\centering
		\includegraphics[scale=0.25]{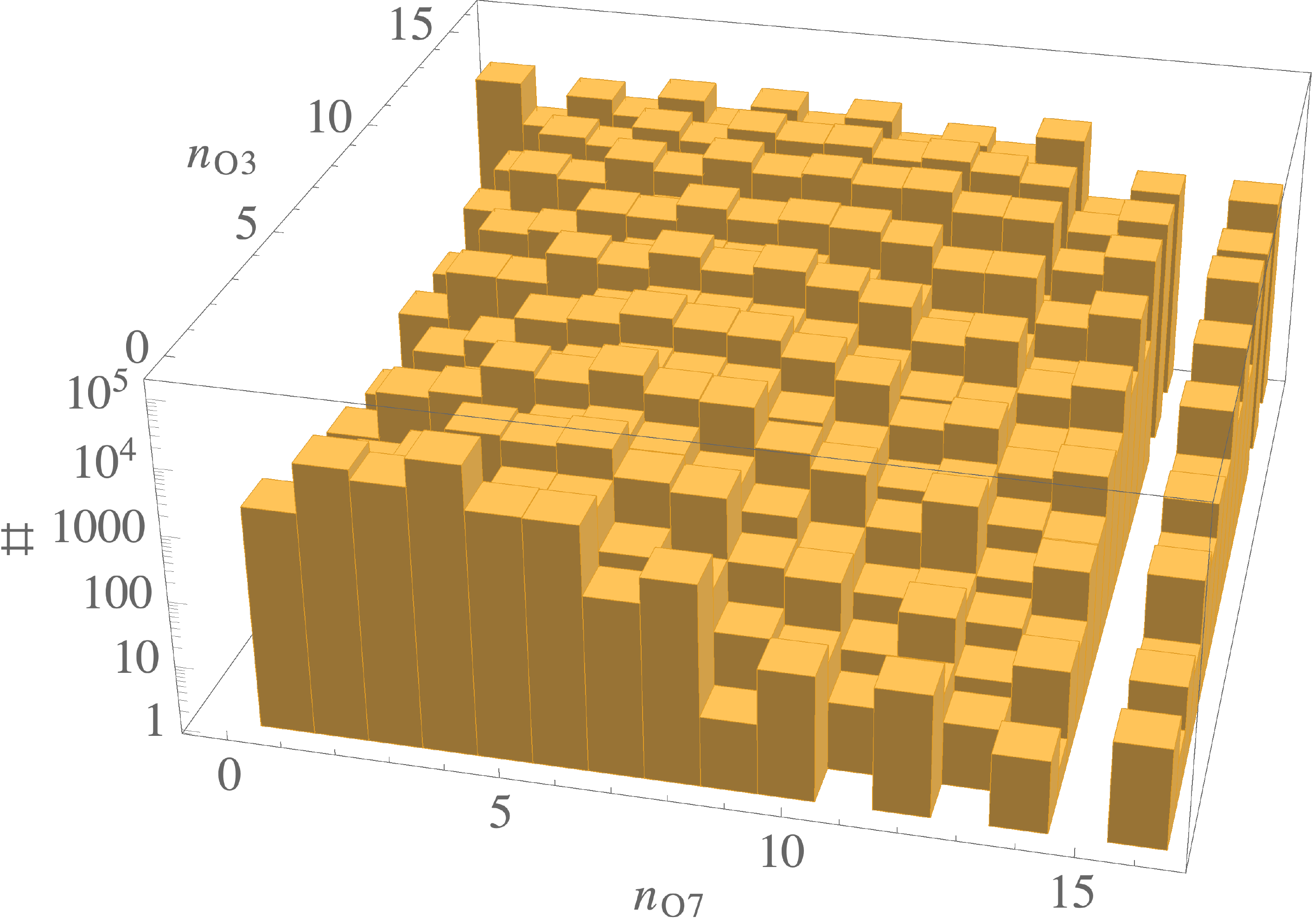}\includegraphics[scale=0.25]{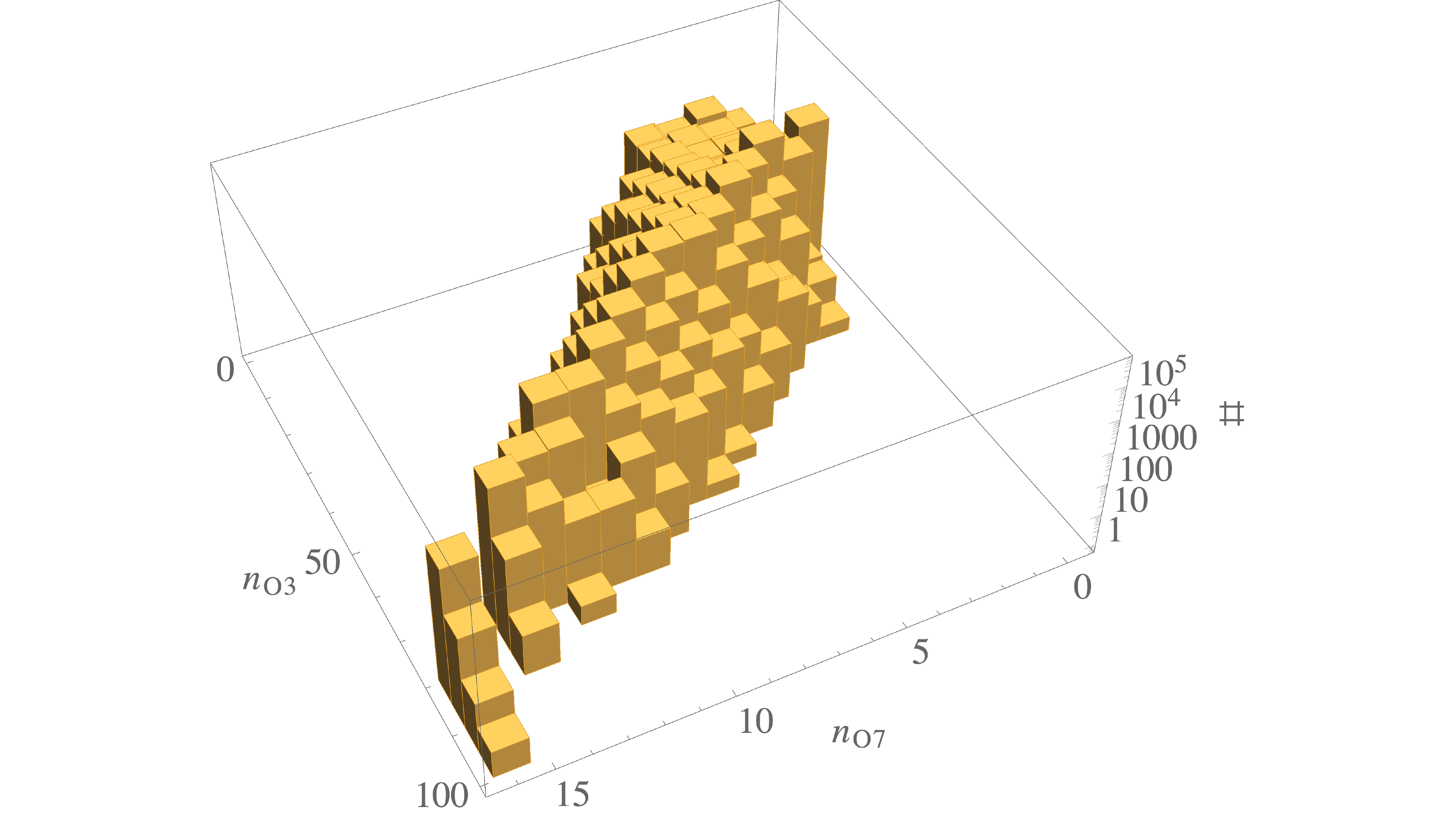}
		\caption{Histograms showing on a logarithmic scale the number of orientifolds producing a given number of O7 planes $n_{O7}$ and O3 planes $n_{O3}$. Left: The minimal number of O3 planes from resolution branches that do not produce extra O3 planes. Right: The maximal number of O3 planes from the resolution branches that produce one extra O3 plane per conifold singularity.}
		\label{fig:nO7nO3histos}
	\end{figure}

	\item The D3-tadpole: The total D3 charge of O3 planes and induced charge on seven-branes is generically negative and can be canceled via the introduction of three-form fluxes. The larger the tadpole (which we define as \textit{minus} the induced D3 charge on O3-planes and 7-branes) the more freedom there is in choosing different three-form fluxes. For all orientifolds, we compute this number for configurations with four D7 branes on top of each O7-plane. We compute the generically much larger D3 tadpole for a generic D7-brane configuration for a subset of the orientifolds of smooth CICYs (see figure~\ref{fig:QD3SO8AllandMaxSmoothHisto}).
	\begin{figure}%[t!]
		\centering
		\includegraphics[scale=0.285]{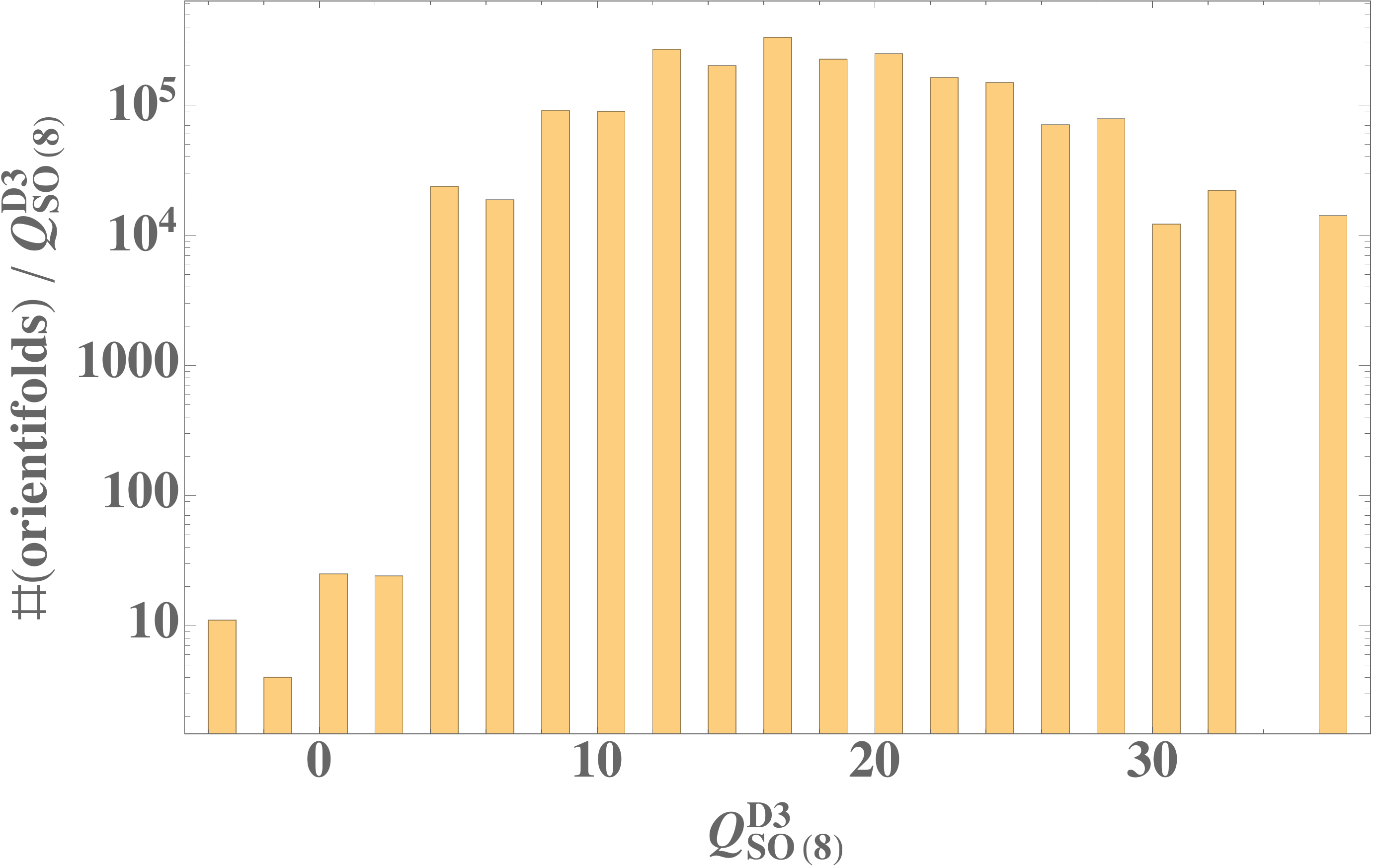}\;\includegraphics[scale=0.3]{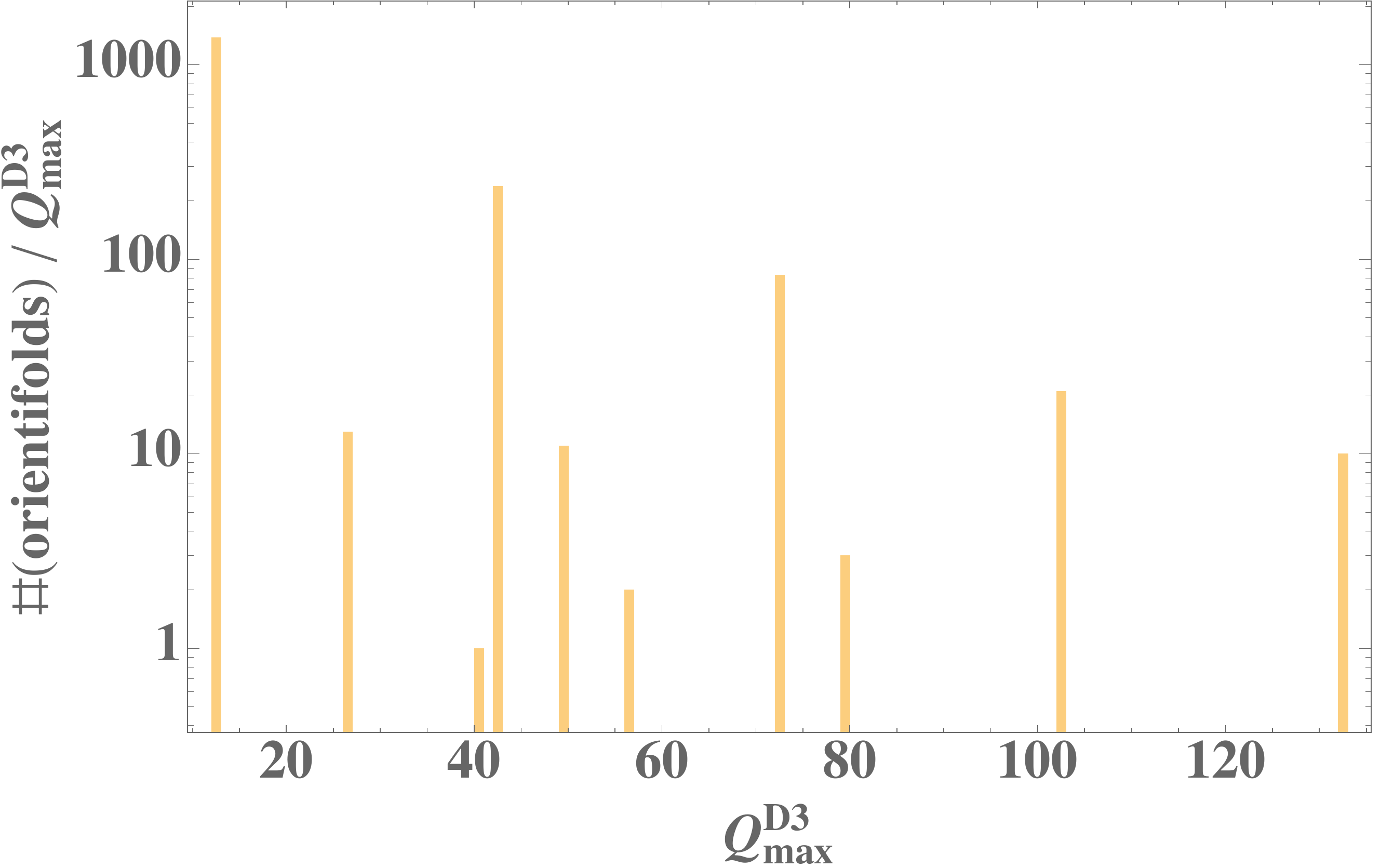}
		\caption{Left: Histogram showing the number of orientifolds with given D3-tadpole $-Q^{D3}_{SO(8)}=\chi_{f}/4$ for local D7-charge cancellation. The tadpole takes \textit{even} values in $[-4,36]$. Right: D3-tadpole for a small set of examples with non-local tadpole cancellation by a generic D7-brane. We have restricted ourselves to the subset of smooth CICYs where the O7 divisor obviously descends from an ambient space divisor. The tadpole takes even values in a significantly larger range $[12,132]$.}
		\label{fig:QD3SO8AllandMaxSmoothHisto}
	\end{figure}
\end{enumerate}

\begin{figure}%[t!]
\centering
\includegraphics[scale=0.23]{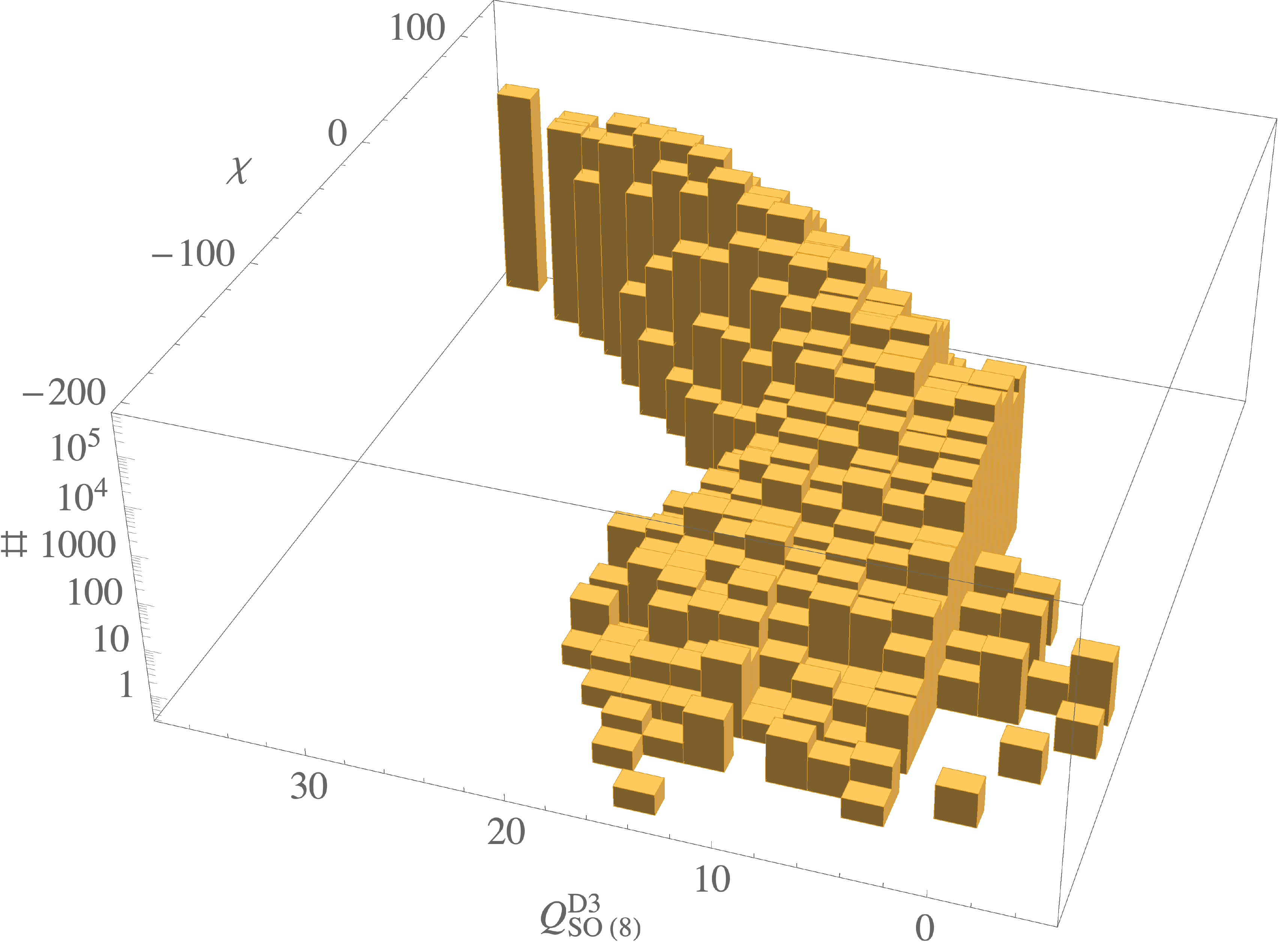}\includegraphics[scale=0.23]{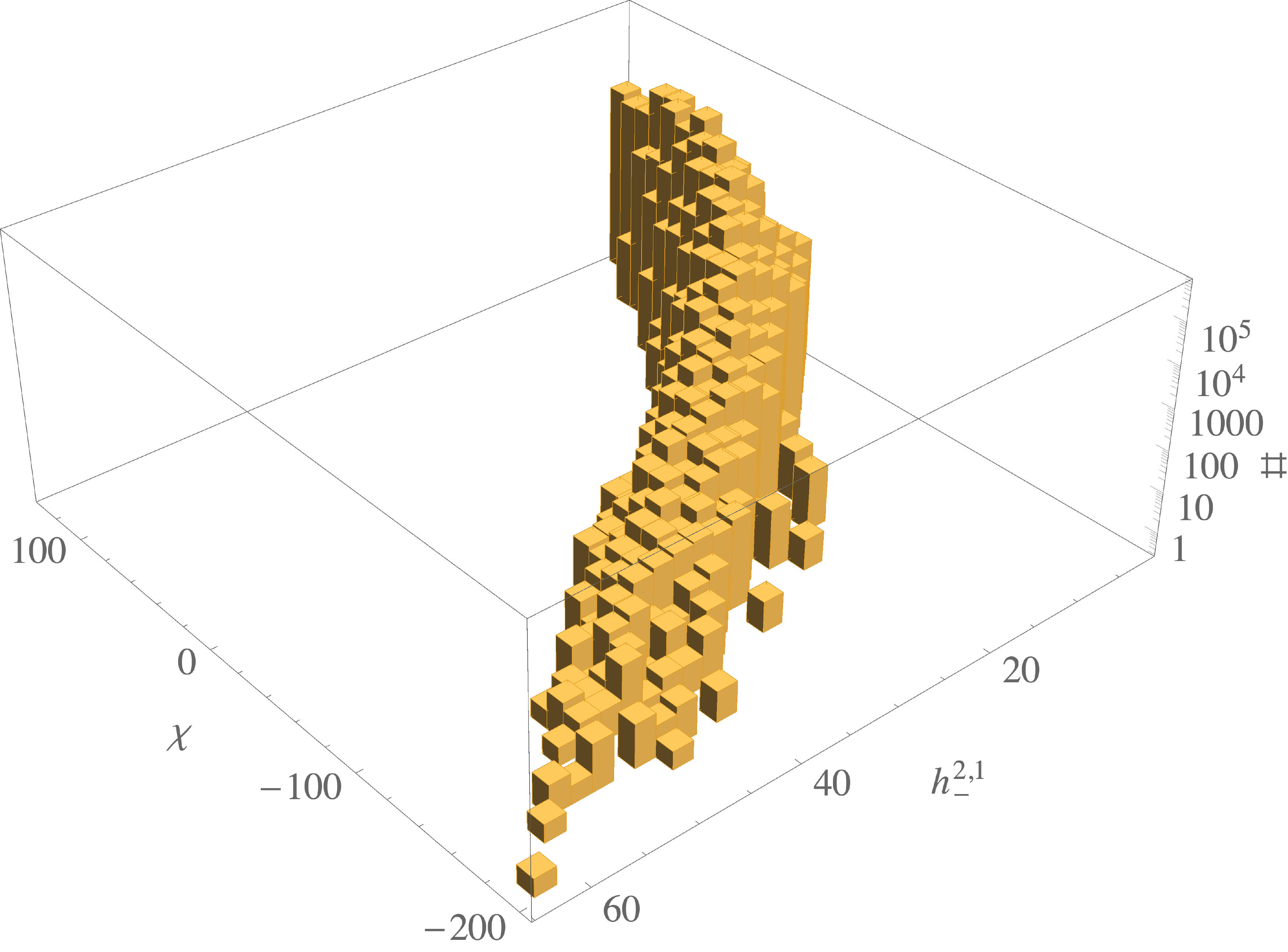}
\caption{Histograms showing on a logarithmic scale the number of orientifolds. Left: Count for each value of $(Q^{D3}_{SO(8)} , \chi)$. Right: Count for each value of $(h^{2,1}_- , \chi)$. The steep cliffs carve out clear islands embedded in large otherwise empty regions.}
\label{fig:QD3chi_h21minusChi_histos}
\end{figure}

Our results indicate strong correlations between the topological quantities that we have computed. More so, our CICY-landscape clearly populates regions in the total parameter space that have pronounced \textit{boundaries} that cannot be explained by the mere finiteness of our sample size. Two more such structures are visible in figure~\ref{fig:QD3chi_h21minusChi_histos}, where we display 3D histograms showing the number of orientifolds for each value of $(Q^{D3}_{SO(8)} , \chi)$ and for each value of $(h^{2,1}_- , \chi)$. The steep cliffs delineate clear islands embedded in otherwise empty regions. It is tempting to speculate that some of these boundaries in fact mark the end of the landscape and the beginning of the swampland. Clearly, it would be interesting to verify or refute this by finding orientifolds in other CY-datasets such as the Kreuzer-Skarke list \cite{Kreuzer:2000xy}.

\section{O3/O7 orientifolds of CICYs}\label{sec::construction}
\subsection{Defining a CICY involution}
We enumerate possible $\mathbb{Z}_2$ actions on complete intersection Calabi-Yau (CICY) manifolds which have a fixed point locus with connected components of complex co-dimension one or three. The CICY manifolds of Candelas et al \cite{Candelas:1987kf} arise as the common vanishing locus of a set of $K$ homogeneous polynomials in an ambient space formed by a product of $r$ projective spaces $\mathbb{P}^{n_1}\times \cdots \times \mathbb{P}^{n_r}$. The (non-unique) starting point for each such manifold is the \textit{configuration matrix}
\begin{equation}
\left[\begin{tabular}{c | c c c}
$\mathbb{P}^{n_1}_{(1)}$ & $m^1_1$ & $\cdots$   & $m_1^K$\\
$\vdots$ & $\vdots$ &  & $\vdots$ \\
$\mathbb{P}^{n_r}_{(r)}$ & $m^1_r$ & $\cdots$  & $m_r^K$
\end{tabular}\right]\, ,
\end{equation}
where the entry $m^i_j$ denotes the weight of the $i$-th polynomial under the scaling of the $j$-th $\mathbb{P}^{n_j}_{(j)}$. In order for the resulting manifold to be a complex threefold we need that $\sum_{j=1}^{r}n_j=K+3$ and in order for its first Chern class to be zero we also need to impose that $\sum_{i=1}^Km^i_j=n_j+1$. As the latter requirement determines the $n_j$, given the $m^i_j$, the first column of the configuration matrix is often omitted.

We restrict ourselves to geometric involutions of the CICY manifolds that can be extended to involutions of the ambient space. While our aim is to exhaust all such involutions we cannot exclude the possibility that other CY involutions exist that cannot be extended to the ambient space. 

Up to linear equivalence, an involution of a $\mathbb{P}^{n_1}\times \cdots \times \mathbb{P}^{n_r}$ ambient space acts as a combination of involutions of individual $\mathbb{P}^n$ factors \cite{Braun:2010vc}
\begin{align}
\mathcal{I}_p^n:\quad  &\mathbb{P}^n \longrightarrow \mathbb{P}^n\, ,\quad 
[x_0:...:x_n]\longmapsto [-x_0:...:-x_{p-1}:x_{p}:...:x_n]\, ,
\end{align}
with fixed locus the disjoint union
\begin{equation}\label{eq:PnFixedLocus}
\mathcal{F}_{p}^n\dot{\cup} \mathcal{F}_{n-p+1}^n\equiv\{x_0=...=x_{p-1}=0\}\dot{\cup}\{x_{p}=...=x_n=0\}\, ,
\end{equation}
and swaps of two $\mathbb{P}^n$ factors
\begin{align}
\mathcal{S}^n: \quad  \mathbb{P}^n\times \mathbb{P}^n \longrightarrow \mathbb{P}^n\times \mathbb{P}^n\, , \quad (x,y)\longmapsto (y,x)\, ,
\end{align}
with fixed point locus the diagonal $\mathbb{P}^n$. Clearly, $\mathcal{I}^n_p$ gives rise to the same $\mathbb{Z}_2$ action as $\mathcal{I}^{n}_{n-p+1}$ because $\mathcal{I}^n_p=-\mathcal{I}^{n}_{n-p+1}$ as matrices acting on the projective coordinates, and $\mathcal{I}^n_0\sim \mathcal{I}^n_{n+1}$ is the trivial involution. Near the locus $\mathcal{F}_{p}^n$ we may use $x_0,...,x_{p-1}$ as local transverse coordinates that are inverted by $\mathcal{I}^n_p$ while near $\mathcal{F}_{n-p+1}^n$ it we may use $x_p,...,x_n$ as local coordinates that are inverted by $\mathcal{I}^n_{n-p+1}$. In particular, a section of some line bundle may be $\mathbb{Z}_2$ symmetric in a local frame around the first fixed point locus and anti-symmetric in another local frame around the second fixed locus.

We encode the ambient space involution into a pair $\left(\vec{\mathcal{I}},\vec{\mathcal{S}}\right)$, with vectors $\vec{\mathcal{I}}=(p_1,...,p_r)^T\simeq \prod_i \mathcal{I}^{n_i}_{p_i}$, and $\vec{\mathcal{S}}=((i_1,j_1),...)^T$ indicating the swaps $((\mathbb{P}^{n_{i_1}}_{(i_1)}~\leftrightarrow~ \mathbb{P}^{n_{j_1}}_{(j_1)}),...)$. Without loss of generality we may assume that the involutions $\mathcal{I}_p$ and the swaps $\mathcal{S}$ act on distinct $\mathbb{P}^n$ factors. The $\mathbb{Z}_2$ generator $g=\prod_{a}\mathcal{I}_{p^a}^a\prod_i \mathcal{S}^{n_i}$ has an action on the set of polynomials $\vec{f}$ via some matrix representation $\mathcal{R}_f(g)$, i.e.
\begin{equation}
g:\quad \vec{f}\longmapsto \mathcal{R}_f(g)\vec{f}\, ,
\end{equation}
where $\mathcal{R}_f(g)$ is again defined up to exchanging $\mathcal{I}^n_{p}$ with $\mathcal{I}^n_{n-p+1}$.
As the action of an involution does not change the scaling weight of a polynomial, we may rotate the polynomials into each other so that $\mathcal{R}_f\left(g\right)$ is block-diagonal with a number of two-dimensional blocks $\sigma_1=\begin{pmatrix}
0 & 1 \\
1 & 0
\end{pmatrix}$ that exchange polynomials of different scaling weights\footnote{If they are the same, we may rotate them into each other in a way that diagonalizes $\sigma_1\longrightarrow \text{diag}(1,-1)$.} and otherwise diagonal entries $\pm 1$.
The $\sigma_1$ blocks correspond to swaps of columns $\vec{\mathcal{S}}_c=((m_1,n_1),...)$ and the combined swaps of rows and columns must leave the configuration matrix invariant. This is necessary for the ambient space involution to map the CY threefold to itself (but not point-wise). Given an ambient space involution we encode the transformation properties of the polynomials in a pair $(\vec{\mathcal{S}}_c,\vec{\mathcal{P}})$ where the \textit{parity vector} $\vec{\mathcal{P}}$ contains entries $+1$ for each polynomial that is either swapped with another one or mapped to itself, and entries $-1$ for each polynomial that is mapped to minus itself. A candidate orientifold of a CICY is thus encoded in the quadruple $(\vec{\mathcal{I}},\vec{\mathcal{S}},\vec{\mathcal{S}_c},\vec{\mathcal{P}})$. 

As an example, consider the ambient space involution of
\begin{equation}
\left[\begin{tabular}{c | c c c}
$\mathbb{P}^{1}$ & $0$ & $0$ & $2$ \\
$\mathbb{P}^{1}$ & $0$ & $2$ & $0$ \\
$\mathbb{P}^{4}$ & $1$ & $2$ & $2$  
\end{tabular}\right]\, ,
\end{equation}
that swaps the two $\mathbb{P}^1$ factors, and inverts two projective coordinates of $\mathbb{P}^4$. The complex structure moduli are assumed to be adjusted so that the second and third two polynomials are exchanged under the ambient space involution, and the first is mapped to minus itself. This is encoded in $\vec{\mathcal{I}}=(0,0,2)^T$, $\vec{\mathcal{S}}=((1,2))^T$, $\vec{\mathcal{S}_c}=((2,3))^T$ and $\vec{\mathcal{P}}=(+1,+1,-1)^T$.

The number of swapped $\mathbb{P}^n$ factors is equal to $h^{1,1}_-$ of the ambient space. We restrict ourselves to configuration matrices that have the property that all the CY divisors descend from ambient space divisors. We follow \cite{Anderson:2017aux} in calling such embeddings \textit{favourable}.\footnote{For favourable embeddings, the CY K\"ahler cone contains the one inherited from the ambient space, but it can have additional generators. If the two cones are equivalent, the embedding is called \textit{K\"ahler-favourable}.} This allows us to simply deduce the induced $\mathbb{Z}_2$ action on the divisor classes of the CY from that of the ambient space. For this purpose it is very useful that in ref. \cite{Anderson:2017aux} the original database of ref. \cite{Candelas:1987kf} has been manipulated to bring all but 70 of the 7890 configuration matrices to a favourable form via \textit{ineffective splittings}, and furthermore giving a useful alternative description of the remaining 70 cases. The latter will be dealt with separately in section \ref{sec:non-fav}.

From eq. \eqref{eq:PnFixedLocus} it follows that the ambient space fixed locus contains $2^{ N_{\mathcal{I}} }$ disjoint components of different dimension, where $N_{\mathcal{I}}$ denotes the number of non-trivial $\mathbb{P}^n$ involutions contained in the ambient space involution. We will tabulate a given ambient space fixed locus by a vector $\vec{q}$ of length $N_{\mathcal{I}}$ containing  $1$'s and $2$'s that indicate which of the two fixed loci in each involuted $\mathbb{P}^n$ factor is to be chosen. For a given ambient space involution and action on the polynomial vector, for each connected component $\mathcal{F}_0$ of the ambient space fixed locus we may always gauge fix the redundancy $\mathcal{I}^n_p\simeq \mathcal{I}^{n}_{n-p+1}$ to find a tuple $(\vec{\mathcal{I}_0},\vec{\mathcal{P}_0})$ such that $\vec{q}=\vec{1}$ corresponds to $\mathcal{F}_0$. We will denote this choice of gauge as the \textit{frame adapted to} $\mathcal{F}_0$. In such a frame, the coordinates transverse to $\mathcal{F}_0$ are inverted, while the action on the projective coordinates that parameterize $\mathcal{F}_0$ is trivial. Given a frame adapted to a fixed locus $\mathcal{F}$, we call $\mathcal{F}$ the \textit{canonical fixed locus}. 

A connected component of the ambient space fixed locus of some given co-dimension $l$ generically descends to a CY fixed locus of co-dimension $l-k_-\geq 0$ where $k_-$ is the dimension of the $(-1)$ eigenspace of the matrix $\mathcal{R}_f(g)$, in a frame adapted to the fixed point locus in question. If $l-k_-< 0$ it does not intersect the CY. $k_-$ is given by the number of $-1$'s in the parity vector $\vec{\mathcal{P}}$ plus the number of pairs of swapped polynomials in $\vec{\mathcal{S}_c}$. This is because the $\mathbb{Z}_2$-odd combinations of polynomials will vanish identically on the ambient space fixed locus.\footnote{Locally, around the fixed locus, we can combine any pair of swapped polynomials $f\leftrightarrow g$ into a $\mathbb{Z}_2$-even function $f+g$ and an odd one $f-g$ although these combinations do not make sense globally when $f$ and $g$ are sections of different line bundles.} Thus, their intersection with the CY is determined by $k_-$ fewer constraints than one would naively expect. For a consistent O3/O7 orientifold we need that each connected component of the ambient space fixed locus descends to a sub-variety of the CY of co-dimension one or three, or completely misses the CY locus. 

\subsection{Singularities at co-dimension one}\label{sec::codim1_sing}
In this paper we will not consider CY three-folds at a locus in their moduli space where they develop singularities of co-dimension one. However, many ambient space involutions whose ambient space fixed point locus intersects the CY at co-dimension one and three feature such singularities. They may arise as follows. Consider a frame adapted to an ambient space fixed locus $\mathcal{F}_0$ of co-dimension $l$ that intersects the CY at co-dimension one. Then, $p\equiv l-1$ polynomials are anti-symmetric around $\mathcal{F}_0$. If any subset of $I\leq p$ anti-symmetric polynomials $f^i_-$ depends non-trivially on $I'\leq I$ of the normal coordinates, we can write
\begin{equation}
f^i_-=\sum_{a=1}^{I'}c^i_a x^a_{\perp}\, ,\quad i=1,...,I\, .
\end{equation}
If $I'\leq I$ the solution set has a component given by setting $x^a_{\perp}=0$ that descends to a variety of dimension three or bigger once intersected with the symmetric polynomials. Such varieties either have components of dimension bigger than three, or contain multiple reducible components of dimension three that generically intersect each other at co-dimension one. Thus, in order to avoid singularities of co-dimension one we should require that each set of $I$ anti-symmetric polynomials depends non-trivially an at least $I'+1$ transverse coordinates. 

With a na\"ive brute-force scan over all subsets of anti-symmetric polynomials we would not be able to complete the orientifold-scan as the computation time grows too quickly with $h^{1,1}$. We note, however, that solving this problem can be mapped to the following different problem: We may define a matrix $A_{\alpha i}$ where each column corresponds to one of the $p$ anti-symmetric polynomials and each row corresponds to one of the $l\equiv p+1$ normal coordinates. We set $A_{\alpha i}=1$ if the i-th anti-symmetric polynomial depends non-trivially on the $\alpha$'th normal coordinate, and zero otherwise. \textit{If} it is possible to find a \textit{vanishing} sub-matrix of dimension $I''\times I$ with at least one column, i.e. $I\geq 1$, and $I+I''\geq l$, then the set of $I$ polynomials depends non-trivially on $I'\equiv l-I''\leq I$ coordinates. In this case, according to what we have said above, there are singularities of co-dimension one. Here is an example: Consider the CICY with number 7734, given by the configuration matrix
\begin{equation}
\left[\begin{tabular}{c | c c c c}
$\mathbb{P}^{1}$ & $2$ & $0$ & $0$ & $0$ \\
$\mathbb{P}^{3}$ & $1$ & $1$ & $1$ & $1$ \\
$\mathbb{P}^{3}$ & $0$ & $2$ & $1$ & $1$  
\end{tabular}\right]\, ,
\end{equation}
and an orientifold specified by the involution $\mathcal{I}=(1,0,2)^T$, no row or column swaps, and parity $\mathcal{P}=(1,-1,1,-1)$. We consider its canonical fixed locus. There are two anti-symmetric polynomials, and three normal coordinates (one normal to the point $x_0=0$ in $\mathbb{P}^1$, and two normal to $\mathbb{P}^1\subset \mathbb{P}^3$). The dependence of the two anti-symmetric polynomials on these normal coordinates is encoded in the matrix
\begin{equation}
A=\begin{pmatrix}
0 & 0\\
1 & 1\\
1 & 1
\end{pmatrix}\, .
\end{equation}
Clearly, there is a vanishing sub-matrix of dimension $(1,2)$, so $3=I+I''\geq l=3$. Thus, the two anti-symmetric polynomials depend non-trivially only on two transverse coordinates and there are co-dimension one singularities in this $\mathbb{Z}_2$-symmetric CY. For this example, this is obvious, but an efficient search for null-submatrices in larger matrices is not as straightforward.

We solve this problem as follows: To $A$ we can associate a \textit{graph} $\Gamma$ as follows. There are $p+l$ vertices, one for each row and one for each column. We connect with an undirected edge each pair of rows and each pair of columns, and connect each pair of row and column if and only if the corresponding matrix element vanishes. In our example, this produces the following graph,
\begin{equation}
\begin{tabular}{m{1cm} m{3cm}}
$\Gamma=$ & \includegraphics[keepaspectratio,width=3cm,trim={3cm 16cm 3cm 3cm},clip]{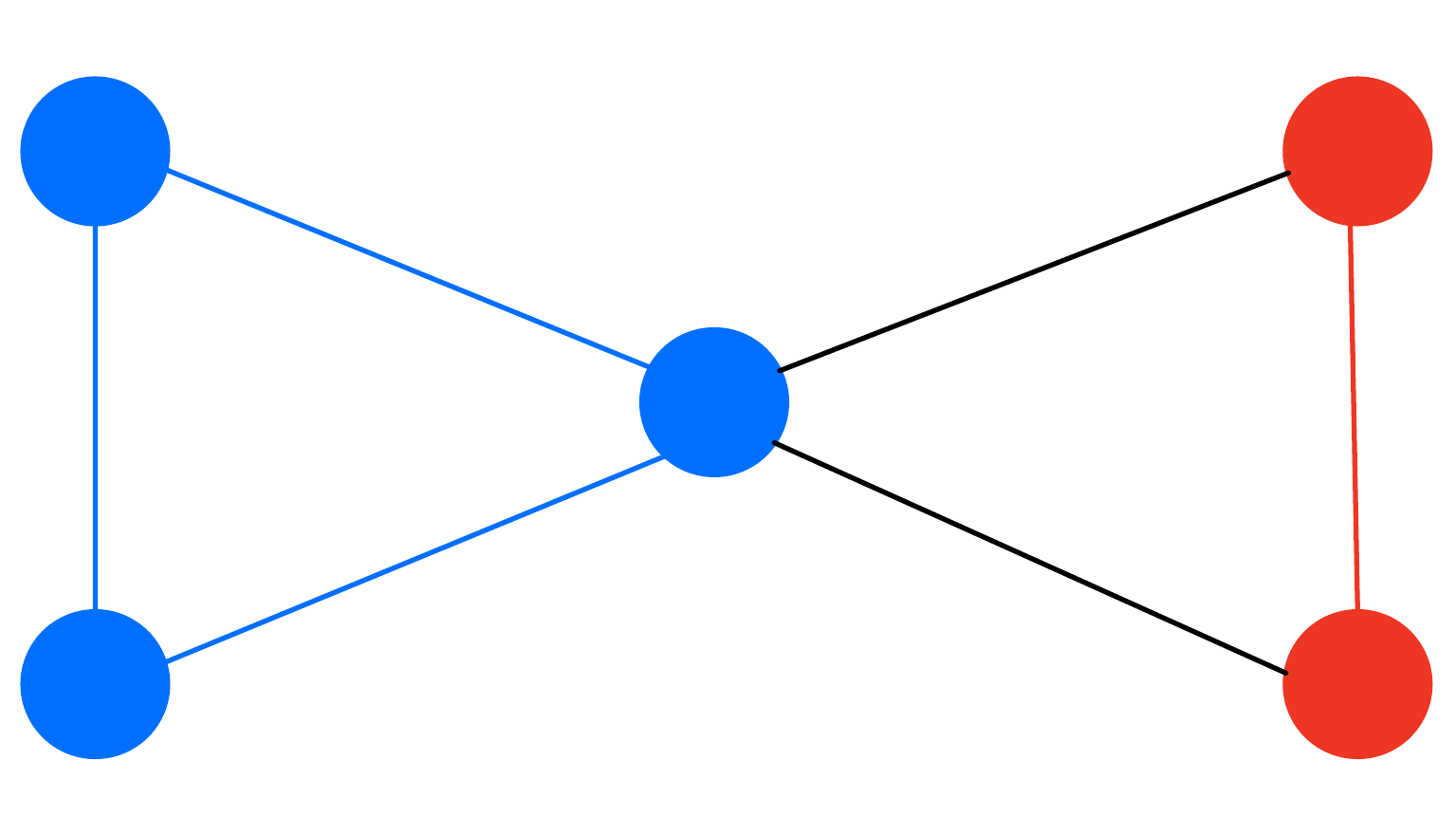}
\end{tabular}\, ,
\end{equation}
where blue vertices correspond to rows and red vertices correspond to columns. The middle vertex correspond to the first row and is thus connected by undirected edges to both column vertices.

Finding the largest possible vanishing submatrix (with size defined as $I+I''$) corresponds to finding the largest \textit{clique} $\Gamma_c$ (i.e. complete subgraph) that contains at least one of the column vertices. In our example this is given by the maximal clique,
\begin{equation}
\begin{tabular}{m{1cm} m{1.5cm}}
$\Gamma_c=$ & \includegraphics[keepaspectratio,width=1.5cm,trim={3cm 13cm 4.5cm 2.5cm},clip]{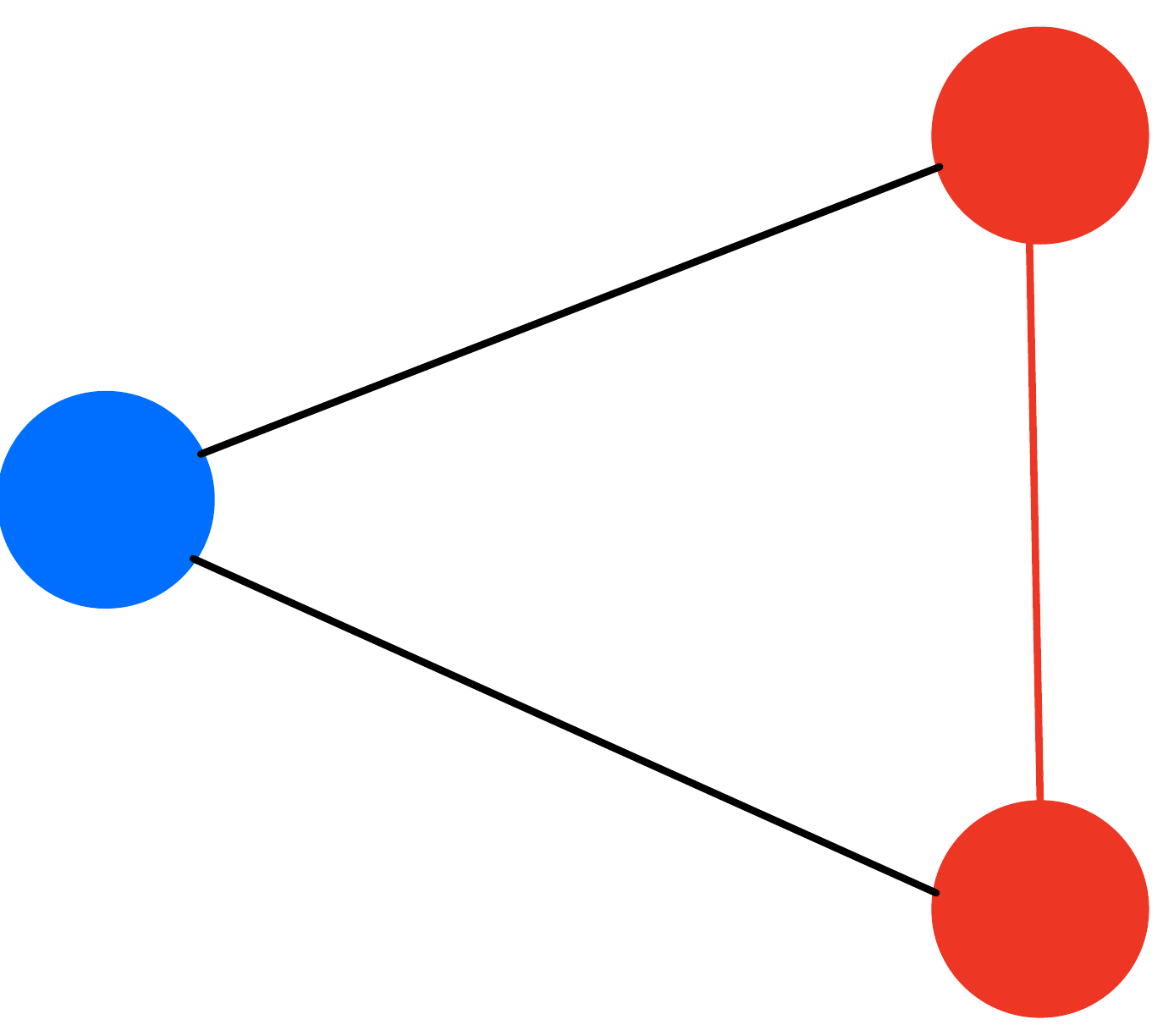}
\end{tabular}\, ,
\end{equation}
with clique number equal to $3$.

The maximal clique problem is non-trivial in general (see e.g. \cite{Aaronson:2005qu}), though easily implemented in \textit{Mathematica}. This step of the computation dominates the overall computational cost at large $h^{1,1}$.
\subsection{Singularities at co-dimension three}\label{sec::codim3_sing}
Having avoided singularities at co-dimension one, most $\mathbb{Z}_2$ symmetric CYs still contain singularities of co-dimension three, generically conifold singularities, that cannot be \textit{deformed} in a $\mathbb{Z}_2$ preserving way, but instead can be \textit{resolved}.\footnote{We thank Fabian R\"uhle for a comment that led us to drop the requirement that the orientifolds are smooth as orientifolds of the original CICY manifolds.} These singularities always reside on top of O7 planes which wrap divisor classes that have no correspondent in the deformed CY, but instead correspond to the new divisor classes that arise upon resolving the conifolds (we will call them \textit{resolution divisors}). As the orientifold projection eliminates the deformation (Coulomb) branch in complex structure moduli space, we denote these singularities \textit{frozen conifold singularities}. 

The generic appearance of this phenomenon is seen as follows.\footnote{For the basics of applied algebraic geometry, see e.g. the introductory chapters of \cite{Hori:2003ic}.} We consider a component $\mathcal{F}$ of the ambient space fixed locus of co-dimension $l$. In order for this to descend to a divisor inside the CY the dimension of the $\mathbb{Z}_2$-odd eigenspace of $\mathcal{R}_f(g)$ must be $l-1$. Then, the intersection of $\mathcal{F}$ with the divisors $D^+_i$ associated with the $\mathbb{Z}_2$ symmetric polynomials is a surface $\mathcal{B}$, but its divisor class need not descend from an ambient space divisor class, unless of course $l=1$. For now, let us assume for simplicity that the anti-symmetric polynomials $f_-^k$ all take values in the same divisor line bundle $\mathcal{O}(E)$. Then, the pull back of their differentials to the fixed locus $\mathcal{F}_l$ take values in the co-normal bundle $N^*\mathcal{F}$ tensored with $\mathcal{O}(E)$ because the dependence on the local coordinates of the fixed locus $\mathcal{F}$ dies out near $\mathcal{F}$. The system of differentials of the $l-1$ anti-symmetric sections degenerates at a co-dimension two locus along $\mathcal{F}$ that is Poincar\'e dual to the Chern class $c_2(N^*\mathcal{F}\otimes \mathcal{O}(E))$. Thus, the number of conifold singularities along the CY divisor is given by
\begin{equation}\label{eq:pre_formula_nconifold}
n_{cf}^{\mathcal{B}}=\int_{\mathcal{B}}c_2(N^*\mathcal{F}\otimes \mathcal{O}(E))\, .
\end{equation}
This is straightforward to evaluate because
\begin{equation}
N\mathcal{F}=\bigoplus_{\mathcal{I}_p^n}\mathcal{O}_{\mathbb{P}^n}(1)^{p} \oplus \bigoplus_{\mathcal{S}^m}\frac{\mathcal{O}_{\mathbb{P}^m_d}(1)^{m+1}}{\mathbb{C}^*}\, ,
\end{equation}
where the first factor comes because the fixed locus of an involution is given by fixing $p$ projective coordinates and the second comes about because the normal bundle of the diagonal $\mathbb{P}^n_d$ is the same as its tangent bundle, and the latter is isomorphic to $\mathcal{O}_{\mathbb{P}^n_d}(1)^{n+1}/\mathbb{C}^*$ as is seen from the Euler sequence. Furthermore, using $ch(E\otimes F)=ch(E)ch(F)$, one shows that
\begin{equation}
c_2(N^*\mathcal{F}\otimes \mathcal{O}(E))=c_2(N\mathcal{F})-(l-1)E\cdot c_1(N\mathcal{F})+\frac{l(l-1)}{2}E^2\, .
\end{equation}
This needs to be generalized to the case where the antisymmetric polynomials are actually sections of $l-1$ distinct line bundles $\mathcal{O}(E_i)$. The appropriate generalization of eq. \eqref{eq:pre_formula_nconifold} is
\begin{equation}\label{eq:masterformula_nconifold}
n_{cf}^{\mathcal{B}}=\int_{\mathcal{B}}\left\{c_2(N\mathcal{F})-\sum_{i}^{l-1}E_i\cdot c_1(N\mathcal{F})+\sum_{i\leq j}^{l-1}E_i\cdot E_j \right\}\, .
\end{equation}
A configuration matrix description for the fixed divisor $\mathcal{B}$ is easily determined (and can be found \href{https://www.desy.de/~westphal/orientifold_webpage/cicy_orientifolds.html}{here}):
\begin{enumerate}
	\item For each involution $\mathcal{I}^n_p$ one replaces $\mathbb{P}^n\longrightarrow \mathbb{P}^{n-p}$ in the configuration matrix.
	\item For each $\mathbb{P}^n$ swap one sums up the two corresponding rows.
	\item All columns associated with anti-symmetric polynomials are omitted.
	\item For each pair of swapped columns, either one of the two is omitted.
\end{enumerate}
The divisor line bundles $\mathcal{O}(E_i)$ are those associated with the omitted columns.\footnote{Note that the anti-symmetric part of a pair of swapped polynomials makes sense only locally near the fixed divisors. They make sense as sections of line bundles on $\mathcal{B}$ associated with either of the swapped columns, which cannot be obtained as pullbacks of ambient space line bundles.} This description also allows one to compute the triple intersection numbers between the resolution divisors wrapped by O7 planes and the bulk ones, by intersecting with the appropriate ambient space divisor classes. The triple intersection numbers that involve two or three resolution divisors wrapped by distinct O7 planes vanish identically, because the fixed locus of eq. \eqref{eq:PnFixedLocus} is a disjoint union. In general there can be further resolution divisors that are not wrapped by O7 planes. We leave the computation of the triple intersection numbers that involve these, as well as the dimension of the resolution branch for future work.\footnote{Its dimension is of course bounded from below by the number of O7 planes containing conifold singularities.} Knowing the dimension of the resolution branch would allow one to determine the pair $(h^{2,1}_+,h^{1,1}_+)$, i.e. the number of vector multiplets and K\"ahler moduli for each orientifold. 

Let us explain why the frozen conifold singularities can always be resolved in a way that maintains the CY condition, producing an O3/O7 orientifold of a different \textit{smooth} CY threefold. Before applying the orientifold projection, in the local non-compact conifold geometry it is always possible to deform or resolve the singularity \cite{Candelas:1989js}. Globally, however, there is generically an obstruction against the resolution in the form of a D-term potential for Strominger's hypermultiplets \cite{Strominger:1995cz,Greene:1995hu}. It admits a flat resolution (Higgs) branch if and only if at the singular locus in moduli space a collection of at least two three-cycles $\{A_i\}$ has shrunken that satisfy at least a single homology relation $\sum_i [A_i]=0$, i.e. there exists a four-chain $\tilde{\Sigma}_4$ s.t. $\sum_i A_i=\del \tilde{\Sigma}_4$ \cite{Greene:1995hu}. Upon resolving, this four-chain turns into a divisor that intersects the exceptional $\mathbb{P}^1$ transversally. Now consider one of the O7 divisors of our setting that passes through at least one conifold singularity. Locally, around one of the singularities it looks precisely like the resolution divisor $\Sigma_4$ (for details see the discussion in the next section \ref{sec:conifold}), so upon deforming it turns into a four-chain with boundary containing the deformed three-sphere $A$ at the tip of the conifold. If the boundary of this four-chain were to contain only this single connected component $A$, we could compute its volume by integrating the holomorphic threeform $\Omega$, i.e. $\text{Vol}(A)\sim |\int_{A}\Omega|=|\int_{\tilde{\Sigma_4}}d\Omega|$,
which vanishes because $\Omega$ is closed, in contradiction to our assumption that we had deformed the singularity. Therefore, there must exist a set of at least two shrunken three-spheres that form the boundary of the four-chain $\tilde{\Sigma_4}$ and so a resolution is possible, as claimed. As argued in the next section, the orientifolding does not project out the resolution branch.

\subsection{Conifolds on O7 planes \& their resolution branches}\label{sec:conifold}
We have stated that many O3/O7 orientifolds contain O7 planes that pass through conifold singularities in the singular CY threefold that can be resolved, but not deformed. Here we would describe the local description of the orientifold in the conifold geometry. Locally, we may write the deformed conifold as the locus \cite{Candelas:1989js}
\begin{equation}\label{eq:conifold}
\det Z=0\, ,\quad Z\equiv\begin{pmatrix}
x & u\\
v & y
\end{pmatrix}=\epsilon\, ,\quad (x,y,u,v)\in \mathbb{C}^4\, ,
\end{equation}
where $\epsilon$ is the deformation parameter. Consider now the involution
\begin{equation}
x\longrightarrow -x\, ,\quad v\longrightarrow -v\, .
\end{equation}
In order for this to be a symmetry of the conifold, we have to take the singular limit $\epsilon\longrightarrow 0$. Only then, the defining eq. \eqref{eq:conifold} transforms homogeneously under the $\mathbb{Z}_2$ action, and the deformation of the conifold is projected out by the orientifold. We are left with an O7 plane residing on the locus $x=v=0$ that passes through the singular point $x=y=u=v=0$.  Although it contains the conifold singularity, it wraps a smooth surface $\mathbb{C}^2\subset \mathbb{C}^4$. In other words, at the singular point, the tangent bundle to the fixed surface inside the CY threefold is regular, while the normal bundle of its embedding into the conifold degenerates.

The symmetry group of the singular conifold is $SU(2)\times SU(2)\times U(1)_R$ and a group element $(L,R,e^{i\phi})$ acts as
\begin{equation}
Z\longrightarrow e^{i\phi} LZR^{\dagger}\, ,
\end{equation}
on the conifold coordinates. The geometric involution corresponds to the group element $(\id,-i\sigma_3, e^{i\pi/2})$, and the deformation modulus is projected out by the orientifolding due to its R-charge $r=2$, i.e. it has a spurious $\mathbb{Z}_2$ transformation $\epsilon\longmapsto-\epsilon$. 

Locally, there are two ways to resolve the conifold with an exceptional $\mathbb{P}^1$ by replacing the defining equation \eqref{eq:conifold} either by
\begin{equation}
A:\quad   Z\cdot\begin{pmatrix}
\alpha\\
\beta
\end{pmatrix}=0\, ,\quad \text{or}\quad B:\quad  Z^T\cdot\begin{pmatrix}
\alpha\\
\beta
\end{pmatrix}=0\, ,\quad  [\alpha,\beta]\in \mathbb{P}^1\, ,
\end{equation}
which we will refer to as the A-type and B-type resolution respectively. Locally, and before orientifolding, these two resolution look the same and are related by a flop transition (see e.g. \textit{Example 7.6.4} of \cite{Hori:2003ic}).

The would-be singular point $Z=0$ is replaced by an exceptional $\mathbb{P}^1$ parameterized by the two projective coordinates $[\alpha,\beta]$. For the A-type resolution, the resolved conifold is $\mathbb{Z}_2$ invariant if the $\mathbb{P}^1$ coordinates transform as
\begin{equation}
[\alpha,\beta]\longmapsto [-\alpha,\beta]=[\alpha,-\beta]\, .
\end{equation}
Thus, on the $\mathbb{P}^1$, there are now \textit{two} fixed points $[1,0]$ and $[0,1]$. The former is a point on the fixed divisor $\sim \mathbb{C}^2$, while the latter is an isolated fixed point. Thus, upon resolving the conifold, we are left with an O7 plane wrapping $\mathbb{C}^2$, transversally intersecting the exceptional $\mathbb{P}^1$, and a disjoint O3 plane. Thus, one may think of the singular orientifold of the conifold as containing an O3 plane collapsed onto an O7 plane. For the B-type resolution, the $\mathbb{P}^1$ coordinates are left invariant by the $\mathbb{Z}_2$ action, and the fixed locus is given by
\begin{equation}
x=v=\alpha u+\beta y=0\, ,
\end{equation}
which is $\mathbb{C}^2$ parameterized by $(u,y)$, blown up at the origin. In a compact setting, $\mathbb{C}^2$ would be replaced by the surface $\mathcal{B}$ as described in the previous section, so the O7 wraps a surface $\tilde{\mathcal{B}}$ defined as $\mathcal{B}$ blown up at $n^{cf}_{\mathcal{B}}$ points. Its Euler characteristic is $\chi(\tilde{\mathcal{B}})=\chi(\mathcal{B})+n^{cf}_{\mathcal{B}}$. The A-type and B-type resolutions are related by a flop transition, under which an SO$(8)$ seven-brane stack eats up an O3 plane, while changing its topology in a way that preserves the D3 tadpole. This is similar to a transition through orbifold singularities described in \cite{Denef:2005mm}. 

Finally, on both resolution branches, we have an $\mathcal{N}=2$ hypermultiplet with bosonic components furnished by the resolution modulus, $B_2$ and $C_2$ integrated over the exceptional $\mathbb{P}^1$, and $C_4$ integrated over the divisor transversally intersecting the exceptional $\mathbb{P}^1$. Away from the tip of the resolved conifold $B_2$ and $C_2$ are proportional to the harmonic two-form of $T^{1,1}$ while $C_4$ is proportional to $dr\wedge \omega_3$ where $r$ is the radial coordinate and $\omega_3$ is the harmonic three-form of $T^{1,1}$. As these forms are invariant under the global symmetry group of the conifold, the geometric $\mathbb{Z}_2$ action leaves them invariant as well. Since $C_2$ and $B_2$ are intrinsically odd under the orientifold action, the associated axions are projected out, and we are left with an $\mathcal{N}=1$ chiral multiplet formed by the resolution modulus and the $C_4$ axion contributing to the orientifold even Hodge number $h^{1,1}_+$. As the orientifold-odd Hodge number $h^{1,1}_-$ does not receive contributions from the resolution moduli, we may infer it from geometrical data of the deformed side CY. 

\subsection{The algorithm}
Having laid out the prerequisites, we now explain how precisely we enumerate the set of orientifolds. The \textit{Mathematica} notebook used to execute the algorithm that we now describe can be found \href{https://www.desy.de/~westphal/orientifold_webpage/cicy_orientifolds.html}{here}.
\begin{enumerate}
	\item First, we exhaust all possible choices of pairs of swaps of rows $\vec{\mathcal{S}}$ and swaps of columns $\vec{\mathcal{S}_c}$ whose combined action leaves the configuration matrix invariant.
	\item Next, we exhaust all choices of distinct involutions $\vec{\mathcal{I}}$ that do not act on $\mathbb{P}^n$ factors that are swapped.
	\item Then, we exhaust all choices of \textit{distinct} parity vectors $\vec{\mathcal{P}}$.
	\item Finally, we remove the cases with co-dimension one singularities.
\end{enumerate}
(a) can be done using brute-force methods. (b) is trivial as well: For each $\mathbb{P}^n$ factor that is not exchanged with any other $\mathbb{P}^n$ factor, one simply goes through the involutions $(\mathcal{I}_1^n,...,\mathcal{I}_{\left[\frac{n+1}{2}\right]}^n)$. The upper bound on the number of inverted coordinates can be seen as a gauge fixing constraint for the $\mathbb{Z}_2$ gauge equivalences $\mathcal{I}^n_{p}\sim \mathcal{I}^n_{n-p+1}$. This fixes the gauge \textit{except} for $\mathbb{P}^n$ factors with $n$ odd, and involutions $\mathcal{I}^{n}_{\frac{n+1}{2}}$ which are mapped to themselves under the $\mathbb{Z}_2$ actions. These will be referred to as \textit{self-dual rows} and will be relevant in what follows. (c) is done by listing all combinations of $\pm$ entries in $\mathcal{P}$ subject to two gauge fixing constraints: First, the polynomials that are swapped can be chosen even without loss of generality. Second, for every self-dual row there remains an unfixed $\mathbb{Z}_2$ gauge symmetry that exchanges the involutions
\begin{align}
&\{[x_0,...,x_n]\mapsto [x_0,...,x_{\frac{n+1}{2}},-x_{\frac{n+1}{2}+1},...,-x_n]\}\nonumber\\
&\leftrightarrow \{[x_0,...,x_n]\mapsto [-x_0,...,-x_{\frac{n+1}{2}},x_{\frac{n+1}{2}+1},...,x_n]\}\, ,
\end{align}
and the polynomials transform with charge equal to the corresponding configuration matrix entry modulo 2. As an example, consider the CICY 7701,
\begin{equation}
\left[\begin{tabular}{c | c c c c }
$\mathbb{P}^{1}$ & $1$ & $1$ & $0$ & $0$  \\
$\mathbb{P}^{1}$ & $1$ & $0$ & $1$ & $0$  \\
$\mathbb{P}^{1}$ & $1$ & $0$ & $0$ & $1$ \\
$\mathbb{P}^{1}$ & $2$ & $0$ & $0$ & $0$ \\
$\mathbb{P}^{3}$ & $1$ & $1$ & $1$ & $1$ 
\end{tabular}\right]\, ,
\end{equation}
and orientifold involution with $\mathcal{I}=(1,1,1,0,2)$, and no row and column swaps. The first three rows and the last one are self-dual, so we produce a charge matrix
\begin{equation}
\begin{pmatrix}
$1$ & $1$ & $0$ & $0$ \\
 $1$ & $0$ & $1$ & $0$ \\
  $1$ & $0$ & $0$ & $1$ \\
   $1$ & $1$ & $1$ & $1$ 
\end{pmatrix}\, ,
\end{equation}
where each row corresponds to an unfixed $\mathbb{Z}_2$ gauge symmetry, and each column collects the $\mathbb{Z}_2$-charges of one of the four polynomials.

In general, one finds a gauge fixing constraint that fixes a maximal set of $\mathbb{Z}_2$-redundancies by taking the charge matrix, reducing modulo two, applying row-reduction over $\mathbb{Z}_2$, and fixing the polynomials associated with the first non-vanishing entry in each row to be positive. In the above example we get
\begin{equation}
\begin{pmatrix}
$1$ & $1$ & $0$ & $0$ \\
$1$ & $0$ & $1$ & $0$ \\
$1$ & $0$ & $0$ & $1$ \\
$1$ & $1$ & $1$ & $1$ 
\end{pmatrix}\, \overset{\text{row reduction}}{\longrightarrow} \begin{pmatrix}
$1$ & $0$ & $0$ & $1$ \\
$0$ & $1$ & $0$ & $1$ \\
$0$ & $0$ & $1$ & $1$ \\
$0$ & $0$ & $0$ & $0$ 
\end{pmatrix}\, .
\end{equation}
Thus we gauge fix $\mathcal{P}=(+1,+1,+1,\pm 1)$.

Finally, (d) is solved by the graph theory exercise described in \ref{sec::codim1_sing}. Before we proceed, let us comment on the data that we extract for each orientifold:
\begin{itemize}
	\item The splitting $h^{1,1}\longrightarrow (h^{1,1}_+,h^{1,1}_-)$. As we are considering favourable CICY descriptions, and the conifold resolution cycles do not contribute to $h^{1,1}_-$, the value of $h^{1,1}_-$ of the CICY coincides with the one of the ambient space. This, in turn, is given by the number of pairs of $\mathbb{P}^n$ factors that are swapped. As we do not know the dimension of the resolution branch, we cannot determine $h^{1,1}_+$ from this information alone.
	\item The splitting $h^{2,1}\longrightarrow (h^{2,1}_+,h^{2,1}_-)$. We use the Lefschetz fixed point theorem to compute $h^{2,1}_-$ in terms of $h^{1,1}_-$,
	\begin{equation}\label{eq:LefschetzFixedPoint}
	h^{2,1}_-=h^{1,1}_-+\frac{\chi_f-\chi_{CY}}{4}-1\, .
	\end{equation}
	From this, and knowledge of the CY Euler characteristic, the value of $h^{1,1}_+-h^{2,1}_+$ follows,
	but without extra input we do not know how to compute $h^{1,1}_+$ and $h^{2,1}_+$ separately. The Euler characteristic of the fixed point set $\chi_f$ is the sum of the Euler characteristics of the fixed divisors and the number of O3 planes. As explained in sections \ref{sec::codim3_sing} and \ref{sec:conifold}, the Euler characteristics that we use are those of the resolution branches of the conifold singularities. For each fixed divisor $\mathcal{B}$ it is given by 
	\begin{equation}
	\chi_{\mathcal{B}}=\chi_{\mathcal{B}}^0+n^{cf}_{\mathcal{B}}\, ,
	\end{equation}
	where $\chi_{\mathcal{B}}^0$ is the Euler characteristic computed by ignoring contributions from the conifold singularities, and $n^{cf}_D$ is the number of conifold singularities that reside on the divisor. Depending on which resolution branch is chosen, the latter contribution is attributed either to points blown up on $\mathcal{B}$, or additional isolated fixed points hosting O3 planes. $\chi_{CY}$ is obtained by adding twice the number of conifold singularities to the Euler characteristic of the CICY we start with \cite{Candelas:1989ug}. As a consistency check we have computed the resulting value of $h^{2,1}_-$ also by counting monomials in examples where this method can be applied (see e.g. Appendix \ref{app:CY-in-dP}). The results match.
	
	\item For each orientifold, we compute the induced D3 brane charge on the seven branes in a configuration where 4 D7 branes sit on top of the fixed point locus, i.e. the O7 plane, thus canceling the D7 tadpole \textit{locally}. It is given by
	\begin{equation}\label{eq:D3charge0}
	-Q^{D3}_{SO(8)}=\frac{1}{4}\chi_f\, ,
	\end{equation}
	as follows from a straightforward expansion of the $\alpha'$ corrected CS action of D7 branes and O7 planes, and the fact that each O3 plane carries $-\frac{1}{4}$ units of D3 charge. This is the \textit{simplest} in that the monodromy transformations around seven-brane stacks are in the center of Sl$(2,\mathbb{Z})$, such that the axio-dilaton does not run. In the generic situation, the D7 branes are split off the O7 plane and have recombined into a single-component D7 brane wrapping a generic $\mathbb{Z}_2$-invariant divisor in the class $8 [O7]$, subject to the constraint that it intersects the O7 plane only along double curves \cite{Braun:2008ua,Collinucci:2008pf}. The resulting D3 tadpole is \cite{Collinucci:2008pf}
	\begin{equation}
	-Q_{U(1)}=-Q^{D3}_{SO(8)}+7[O7]^3\, ,
	\end{equation}
	matches the flux-less F-theory result, and can be much larger than \eqref{eq:D3charge0}. For the small number of orientifolds where the O7 divisor $[O7]$ descends in an obvious way from an ambient space class, we also compute $Q_{U(1)}$. The resulting distribution is shown in figure \ref{fig:QD3SO8AllandMaxSmoothHisto}.
\end{itemize}
For each orientifold we produce an entry
\begin{equation}
\left\{\vec{\mathcal{I}},  \vec{S},  \vec{S}_c, \vec{\mathcal{P}},  \chi_{CY} ,
	 h^{1,1}_-,  h^{2,1}_-,  \{\{\vec{q}_{O7^1}, \chi_{O7^1}, d_{O7^1}, n^{cf}_{O7^1}\},...\}, 
	 \{\{\vec{q}_{O3^1}, n_{O3^1}\},...\}, \frac{1}{4}\chi_f\right\}\, ,
\end{equation}
where $\vec{q}_{O3/O7^i}$ encodes the ambient space fixed locus from which the i-th set of O3 planes or the i-th O7 plane divisor descends, $(\chi_{O7^i},d_{O7^i},n^{cf}_{O7^i})$ denote the Euler characteristic, degree ($\equiv\int (c_1)^2$) and number of conifold singularities on the i-th O7 plane divisor\footnote{$(\chi_{O7^i},d_{O7^i})$ corresponds to the surface wrapped by the O7 on the A-type resolution branch.}, $n_{O3^i}$ is the number of O3 planes in this set, and $n_{O3}$ denotes the total number of O3 planes. The full list of orientifolds can be found \href{https://www.desy.de/~westphal/orientifold_webpage/cicy_orientifolds.html}{here}.

\subsection{Examples}
\subsubsection{The quintic threefold}
We follow traditional practice in starting with the quintic threefold $[\mathbb{P}^4|5]$, number 7890 of the CICY-list. Clearly we cannot swap two rows so all orientifolds will have $h^{1,1}_-=0$. The first involution is $\mathcal{I}^4_1$ which acts as 
\begin{equation}
\mathcal{I}^4_1:\, \mathbb{P}^4\longrightarrow \mathbb{P}^4\, , \quad [x_0:\cdots :x_4]\longmapsto [-x_0:x_1:\cdots :x_4]\, ,
\end{equation}
or shorthand $\vec{\mathcal{I}}=(1)$. Its fixed locus is the disjoint union $\{x_0=0\}\cup \{x_1=x_2=x_3=x_4=0\}$ with co-dimension one respectively four in the ambient space. Choosing the polynomial to be symmetric under the involution (i.e. $\vec{\mathcal{P}}=(1)$), the ambient space divisor $x_0=0$ descends to a CY divisor $D$ hosting an O7 plane. Its Euler number is easily computed to be $\chi_D=55$. Near the second fixed locus it is more useful to write the involution as $[x_0:\cdots :x_4]\longmapsto [x_0:-x_1:\cdots :-x_4]$ because we may use the inverted coordinates as local transverse coordinates to the fixed point. As the weight of the polynomial is odd, it is anti-symmetric across the fixed point, so the co-dimension four ambient space locus $\{x_1=x_2=x_3=x_4=0\}$ descends to an isolated fixed point inside the CY hosting an O3 plane. Thus, the D3 tadpole is $(55+1)/4=14$. We compute the value of $h^{2,1}_-$ using the index theorem,
\begin{equation}
h^{2,1}_-=0+\frac{55+1}{4}-\frac{-200}{4}-1=63\, .
\end{equation}
In total, we produce the following entry for this orientifold,
\begin{equation}
\{\{1\}, \{\}, \{\}, \{1\}, -200, 0, 63, \{\{\{1\}, 55, 5, 0\}\}, \{\{\{2\}, 1\}\}, 14\}\, .
\end{equation}
Next, let us consider the involution $\vec{\mathcal{I}}=(2)$, i.e. 
\begin{equation}
\mathcal{I}^4_1:\, \mathbb{P}^4\longrightarrow \mathbb{P}^4\, , \quad [x_0:\cdots :x_4]\longmapsto [-x_0:-x_1:x_2:x_3 :x_4]\, .
\end{equation}
with two ambient space fixed loci of co-dimensions two respectively three. The polynomial must be chosen anti-symmetric across the first locus so that it descends again to a divisor $D$ hosting an O7 plane rather than a locus of co-dimension two. Thus, $\vec{\mathcal{P}}=(-1)$. The second fixed locus descends to a set of $\int_{CY}H^3=\int_{\mathbb{P}^4}5H^4=5$ isolated fixed points hosting $5$ O3 planes. The divisor $D$ is a $\mathbb{P}^2\subset \mathbb{P}^4$ so it has Euler number $\chi_{D}=3$. The normal bundle is isomorphic to $\mathcal{O}_{\mathbb{P}^2}(1)^2$, so $c_1(N\mathcal{F})=2H$ and $c_2(N\mathcal{F})=H^2$. The anti-symmetric polynomial takes values in $\mathcal{O}(E)=\mathcal{O}_{\mathbb{P}^2}(5)$. Thus, we encounter
\begin{equation}
\int_{\mathbb{P}^2}c_2(N\mathcal{F})-c_1(N\mathcal{F})\cdot E+E^2=1-2\cdot 5+5^2=16
\end{equation}
conifold singularities along it. Therefore, the D3 tadpole is $(5+3+16)/4=6$. Moreover,
\begin{equation}
h^{2,1}_-=0+\frac{5+3+16}{4}-\frac{-200+2\cdot 16}{4}-1=47\, .
\end{equation}
The corresponding entry in our list is
\begin{equation}
\{\{2\}, \{\}, \{\}, \{-1\}, -168, 0, 47, \{\{\{1\}, 3, 9, 16\}\}, \{\{\{2\}, 5\}\}, 6\}\, .
\end{equation}
Note that this is the singular limit of a smooth CICY orientifold, obtained via the B-resolution of the conifold singularity.  It is given by an orientifold of the CICY 7885,
\begin{equation}
\left[\begin{tabular}{c | c c }
$\mathbb{P}^{1}$ & $1$ & $1$\\
$\mathbb{P}^{4}$ & $4$ & $1$      
\end{tabular}\right]\, ,
\end{equation}
with entry
\begin{equation}
\{\{0, 2\}, \{\}, \{\}, \{1, -1\}, -168, 0, 47, \{\{\{1\}, 19, -7, 0\}\}, \{\{\{2\}, 5\}\}, 6\}\, .
\end{equation}

\subsubsection{Anti-canonical hypersurface in $\mathbb{P}^2\times \mathbb{P}^2$}
As a second example we would like to consider an anti-canonical hypersurface in $\mathbb{P}^2\times \mathbb{P}^2$, with CICY number 7884. First, we consider $\vec{\mathcal{I}}=(1,0)$, acting on only one of the $\mathbb{P}^2$s. There are two ambient space fixed loci, of co-dimensions one respectively two. The polynomial has to be chosen symmetric across the first locus which then descends to a divisor $D$. One computes $\chi_D=36$. Then, across the second fixed locus, the polynomial is anti-symmetric giving a $\mathbb{P}^2$ divisor with nine conifold singularities along it. The D3 tadpole is $(36+3+9)/4=12$, and $h^{2,1}_-=47$. The entry is
\begin{equation}
\{\{0, 1\}, \{\}, \{\}, \{1\}, -144, 0, 47, \{\{\{1\}, 36, 0, 0\}, \{\{2\}, 3, 9, 9\}\}, \{\}, 
	12\}\, ,
\end{equation}
containing a $\mathbb{P}^2$ divisor with 9 conifold singularities.
The B-resolution leads to the smooth orientifold of CICY number 7875,
\begin{equation}
\left[\begin{tabular}{c | c c }
$\mathbb{P}^{1}$ & $1$ & $1$  \\
$\mathbb{P}^{2}$ & $1$ & $2$  \\
$\mathbb{P}^{2}$ & $0$ & $3$  \\
\end{tabular}\right]\, ,
\end{equation}
with orientifold-entry
\begin{equation}
 \{\{0, 1, 0\}, \{\}, \{\}, \{1, 1\}, -144, 0, 47, \{\{\{1\}, 36, 0, 0\}, \{\{2\}, 12, 0, 0\}\}, 
	\{\}, 12\}\, .
\end{equation}
The $\mathbb{P}^2$ divisor has been blown up at nine points, producing the rational elliptic surface $dP_9$.

We may also consider involuting both factors, i.e. $\vec{\mathcal{I}}=(1,1)^T$. This gives four ambient space fixed loci of co-dimensions $(2,3,3,4)$. The polynomial must be chosen antisymmetric across the first locus so that it descends to a divisor inside the CY. It has the topology of $\mathbb{P}^1\times \mathbb{P}^1$, and along it one finds $13$ conifold singularities. The other fixed loci give rise to $3+3+1=7$ O3 planes. Thus, the D3 tadpole is $(4+13+7)/4=6$, and $h^{2,1}_-=39$, and the orientifold entry reads 
\begin{align}
&\left\{\{1, 1\}, \{\}, \{\}, \{-1\}, -136, 0, 39, \{\{\{1, 1\}, 4, 8, 13\}\}, \right. \nonumber\\
&\left. 
	\{\{\{1, 2\}, 3\}, \{\{2, 1\}, 3\}, \{\{2, 2\}, 1\}\}, 6\right\}\, .
\end{align}
The Euler characteristic of a resolution of the conifold singularities is $\chi_{CY}=\chi_{CY}^0+2n^{cf}=-162+26=-136$. No such CY threefold is contained in the CICY list. 

Finally, we may swap the two $\mathbb{P}^2$ factors giving $h^{1,1}_-=1$. The diagonal $\mathbb{P}^2$ has a normal bundle isomorphic to its tangent bundle so $c(N\mathcal{F})=(1+H_d)^3$. The omitted anti-symmetric polynomial is in $E=3H_1+3H_2$ which pulls back to $6H_d$ so the divisor contains 
\begin{equation}
3H_d^2-3H_d\cdot 6H_d+(6H_d)^2=21
\end{equation}
conifold singularities. Therefore, the D3 tadpole is $(3+21)/4=6$, and $h^{2,1}_-=35$. The orientifold entry reads
\begin{equation}
\{\{0, 0\}, \{\{1, 2\}\}, \{\}, \{-1\}, -120, 1, 36, \{\{\{\}, 3, 9, 21\}\}, \{\}, 6\}\, ,
\end{equation}
containing a $\mathbb{P}^2$ surface hosting 21 conifold singularities.
The B-type resolution of this orientifold gives the CICY with number 7846,
\begin{equation}
\left[\begin{tabular}{c | c c c }
$\mathbb{P}^{2}$ & $2$ & $1$ & $0$  \\
$\mathbb{P}^{2}$ & $2$ & $0$ & $1$  \\
$\mathbb{P}^{2}$ & $1$ & $1$ & $1$  
\end{tabular}\right]\, ,
\end{equation}
with orientifold entry
\begin{equation}
\{\{0, 0, 0\}, \{\{1, 2\}\}, \{\{2, 3\}\}, \{1, 1, 1\}, -120, 1, 
	36, \{\{\{\}, 24, -12, 0\}\}, \{\}, 6\}\, .
\end{equation}
The $\mathbb{P}^2$ has been blown up at 21 points producing a surface which is not nef.

\subsubsection{$[\mathbb{P}^5|4\,2]$}
Finally, we consider the orientifold of $[\mathbb{P}^5|4\,2]$ (CICY number 7889) specified by $\vec{\mathcal{I}}=(2)$, and $\vec{\mathcal{P}}=(1,-1)^T$. One may show that there is a single O7 plane divisor $D$ with $(\chi_D,d_D,n^{cf}_D)=(24,0,4)$, as well as four O3 planes. The orientifold entry is
\begin{equation}
\{\{2\}, \{\}, \{\}, \{1, -1\}, -168, 0, 49, \{\{\{1\}, 24, 0, 4\}\}, \{\{\{2\}, 4\}\}, 8\}\, ,
\end{equation}
containing a K3 surface with 4 conifold singularities.

In this case, the two different resolutions of the conifold lead to different orientifolds of the same CY with CICY number 7886,
\begin{equation}
\left[\begin{tabular}{c | c c c}
$\mathbb{P}^{1}$ & $0$ & $1$ & $1$   \\
$\mathbb{P}^{5}$ & $4$ & $1$ & $1$
\end{tabular}\right]\, .
\end{equation}
The B-type resolution leads to 
\begin{equation}
\{\{0, 2\}, \{\}, \{\}, \{1, 1, -1\}, -168, 0, 49, \{\{\{1\}, 28, -4, 0\}\}, \{\{\{2\}, 4\}\}, 
	8\}\, ,
\end{equation}
where K3 is blown up at four points.
The A-type resolution leads to 
\begin{align}
&\left\{\{1, 2\}, \{\}, \{\}, \{1, 1, 1\}, -168, 0, 49, \right. \nonumber\\ 
&\left. \{\{\{2, 1\}, 24, 0, 0\}\}, 
	\{\{\{1, 1\}, 4\}, \{\{1, 2\}, 4\}\}, 8\right\}\, ,
\end{align}
and four extra O3 planes have been produced.

Note that while the ability to describe at least one of the resolution branches of a given singular CICY orientifold via a smooth orientifold of a different CICY appears quite frequently at low values of $h^{1,1}$, at larger values of $h^{1,1}$ approximately half of the resolved CYs have \textit{positive} Euler characteristic and are therefore \textit{not} contained in the CICY list.
\subsubsection{An example with vanishing D3 tadpole}
It is also possible to find orientifold vacua that are consistent \textit{without} the inclusion of either D3 branes or three-form fluxes. These are precisely those with vanishing overall D3 tadpole. As an example, consider the CICY 7888 and its smooth orientifold specified by
\begin{equation}
\left[\begin{tabular}{c | c c}
$\mathbb{P}^{1}$ & $0$ & $2$ \\
$\mathbb{P}^{4}$ & $4$ & $1$    \\
\end{tabular}\right]\, , \quad \vec{\mathcal{I}}=(0,2)^T\, ,\quad \vec{\mathcal{P}}=(1,-1)^T\, .
\end{equation}
The first ambient space fixed point locus descends to a divisor $D$ of topology $\mathbb{P}^1\times [\mathbb{P}^2|4]$, and $[\mathbb{P}^2|4]$ is a genus two Riemann surface. Its Euler characteristic is $\chi_D=-8$, and its degree is $d=-16$. There are no conifold singularities along it, so the induced D3 charge on the seven branes is $+2$. The second ambient space fixed point locus descends to a set of 8 isolated fixed points in the CY, hosting O3 planes that precisely cancel the positive contribution from the seven branes. The D3 tadpole therefore vanishes in the absence of fluxes and D3 branes. The orientifold entry is
\begin{equation}
\{\{0, 2\}, \{\}, \{\}, \{1, -1\}, -168, 0, 41, \{\{\{1\}, -8, -16, 0\}\}, \{\{\{2\}, 8\}\}, 0\}\, .
\end{equation}

\subsection{The non-favourable CICYs}\label{sec:non-fav}
Now we turn to the remaining CICY manifolds whose complete set of divisors does not descend from the ambient space divisors via intersection with the CY threefold, i.e. they are not favourable with respect to an ambient space that is a product of $\mathbb{P}^n$'s. Out of the 70 configuration matrices that are left non-favourable, a total of 22 matrices are either $T^6$ or $K3\times T^2$. As we are interested in $\mathcal{N}=1$ orientifolds we discard these. The other 48 matrices fall into two classes \cite{Anderson:2017aux},
\begin{enumerate}
	\item 33 matrices describe CY threefolds that can be rewritten as favourable hypersurfaces in an ambient space $\mathcal{B}_1\times \mathcal{B}_2$ where $\mathcal{B}_{1,2}$ are del Pezzo surfaces. It is well known that the del-Pezzo surfaces $\{\mathbb{P}^2,\mathbb{P}^1\times \mathbb{P}^1,dP_1,...,dP_7\}$ can be obtained as a complete intersection manifold \cite{Hubsch:1992nu}. Thus, the anti-canonical hypersurface in each pairing appears in the list of CICYs. 
	\item 15 matrices describe the same manifold, the Schoen manifold \cite{Schoen1988} which can be obtained as an anti-canonical hypersurface in $dP_9\times dP_9$, where $dP_9$ is a rational elliptic surface, i.e. $\mathbb{P}^2$ blown up at nine points. $dP_9$ can also be written as a CI manifold. The anti-canonical hypersurfaces in $\mathcal{B}_1\times dP_9$ with $\mathcal{B}_1$ almost del-Pezzo surface are all equivalent to the Schoen manifold (Theorem 3.1 of \cite{Hubsch:1992nu}). 
\end{enumerate}
\subsubsection{CY hypersurfaces in products of del-Pezzo surfaces}
Let us first focus on (a). The $\mathbb{Z}_2$ involutions of the ambient space come in two classes. I: Combinations of involutions of the two factors, and II: swaps of the two factors whenever $\mathcal{B}_1\simeq \mathcal{B}_2$. 

We start with I. All possible $\mathbb{Z}_2$ involutions of del Pezzo surfaces have been exhibited in ref. \cite{Blumenhagen:2008zz}.\footnote{We exclude $dP_8$. A boring reason for discarding it is that anti-canonical hypersurfaces of $\mathcal{B}\times dP_8$ do not appear in the CICY list. A better reason for doing so is that $\text{dim}\,H^0(\bar{K},dP_8)=2$. Thus, any global section $F$ of the anti-canonical line bundle on $\mathcal{B}\times dP_8$ can be expanded as $F=f_1g^1+f_2g^2$ with $f_i\in H^0(\bar{K},\mathcal{B})$ and $g^i\in H^0(\bar{K},dP_8)$. Then, a general CY hypersurface has conifold singularities at the $d_{\mathcal{B}}\equiv \int_{\mathcal{B}}c_1(T\mathcal{B})^2$ points where $f_1=f_2=g^1=g^2=0$ that cannot be deformed (i.e. Bertini's theorem cannot be applied). They can, however, be resolved by blowing up the single point on $dP_8$ that is Poincaré dual to $c_{1}(TdP_8)^2$. The result is the Schoen manifold constructed in ref. \cite{Schoen1988}.} There exist $28$ pairs $(\mathcal{B},\sigma)$ where $\mathcal{B}$ is one of the ten del Pezzo surfaces and $\sigma$ is a non-trivial involution, see table \ref{tab:dPinvolutions}. First, consider an involution that acts non-trivially only on (say) the first ambient space factor. Its fixed locus inside $\mathcal{B}_1$ is a union of a set of fixed divisors $D^o_i$ and a set of $n_{fp}$ isolated fixed points. Choosing the anti-canonical section to be $\mathbb{Z}_2$ symmetric the $D^o_i$ descend to generically smooth CY O7 divisors $D^o_{i,CY}$ while the $n_{fp}$ isolated fixed points descend to CY O7 divisors $\tilde{D}^o_{j,CY}$ hosting conifold singularities. It is straightforward to show that the former have Euler characteristic
\begin{equation}
\chi_{D^o_{i,CY}}=12\int_{\mathcal{B}_1}D^o_i\cdot c_{1}(\mathcal{B}_1)\, ,
\end{equation}
and vanishing degree. Each isolated fixed point contributes a del Pezzo CY divisor with the topology of $\mathcal{B}_2$, containing a number of conifold singularities equal to the degree of $\mathcal{B}_2$. The D3 tadpole takes a universal value,
\begin{align}
-Q^{D3}&=\frac{1}{4}\sum_i\chi_{D^o_{i,CY}}+\frac{1}{4}n_{fp}\left(\chi_{\mathcal{B}_2}+\int_{\mathcal{B}_2}c_1(\mathcal{B}_2)^2\right)\nonumber\\
&=3\left(\sum_i\int_{\mathcal{B}_1}D^o_i\cdot c_{1}(\mathcal{B}_1)+n_{fp}\right)= 12\, ,
\end{align}
where the last equality is shown to hold by going through the list of involutions one by one.

Second, we consider involutions that act non-trivially on both del Pezzo factors. The ambient space fixed point locus has various different components. One, surfaces $\mathcal{C}^1_i\times \mathcal{C}^2_j$ where the $\mathcal{C}^{1,2}_i$ are fixed curves inside $\mathcal{B}_{1,2}$, and two, curves for each pairing of isolated fixed points in $\mathcal{B}_1$ with fixed curves in $\mathcal{B}_2$ and vice versa, and three, isolated fixed points for each pairings of isolated fixed points in the two ambient space factors. For an O3/O7 orientifold, the anti-canonical section has to be chosen anti-symmetric across the first locus giving rise to a set of CY divisors $D^o_{ij}$ of topology $\mathcal{C}^1_i\times \mathcal{C}^2_j$ with Euler characteristic
\begin{equation}
\chi_{D^o_{ij}}=(\mathcal{C}^1_i)^2(\mathcal{C}^2_j)^2-(\mathcal{C}^1_i)^2(\mathcal{C}^2_j\cdot \bar{K}_{\mathcal{B}_2})-(\mathcal{C}^2_j)^2(\mathcal{C}^1_i\cdot \bar{K}_{\mathcal{B}_1})+(\mathcal{C}^1_i\cdot \bar{K}_{\mathcal{B}_1})(\mathcal{C}^2_j\cdot \bar{K}_{\mathcal{B}_2})\, ,
\end{equation}
containing 
\begin{equation}
n_{cf}=(\mathcal{C}^1_i)^2(\mathcal{C}^2_j)^2-(\mathcal{C}^1_i)^2(\mathcal{C}^2_j\cdot \bar{K}_{\mathcal{B}_2})-(\mathcal{C}^2_j)^2(\mathcal{C}^1_i\cdot \bar{K}_{\mathcal{B}_1})+2(\mathcal{C}^1_i\cdot \bar{K}_{\mathcal{B}_1})(\mathcal{C}^2_j\cdot \bar{K}_{\mathcal{B}_2})
\end{equation}
conifold singularities. All other ambient space fixed loci give rise to a total number of O3 planes computed as
\begin{equation}
n_{O3}=n^1_{fp}n^2_{fp}+n^1_{fp}\sum_j\mathcal{C}^2_j\cdot \bar{K}_{\mathcal{B}_2}+n^2_{fp}\sum_i \mathcal{C}^1_i\cdot \bar{K}_{\mathcal{B}_1}\, .
\end{equation}
In order to avoid singularities of co-dimension smaller than three we must impose additional requirements. Consider an involution $(\sigma_1,\sigma_2)$ acting non-trivially on both del-Pezzo factors, and consider one of the two factors, say $\mathcal{B}_1$. As just explained, if both $\sigma_{1,2}$ have non-trivial fixed curves $(\mathcal{C}^1,\mathcal{C}^2)$ we get a non-trivial fixed divisor $\mathcal{C}^1\times \mathcal{C}^2$ in the CY if we choose the anti-canonical section $F$ to be anti-symmetric across the surface $\mathcal{C}^1\times \mathcal{C}^2\subset \mathcal{B}_1\times \mathcal{B}_2$ defined via the vanishing of two global sections $f^{i}\in \Gamma(\mathcal{O}(\mathcal{C}^{i}))$. Now, pick a point $p\in \mathcal{C}^2\subset \mathcal{B}_2$ and define $G\equiv F(p)$ which is a global anti-canonical section on $\mathcal{B}_1$. As $g$ is anti-symmetric across $\mathcal{C}^1$ under the involution, it vanishes at least linearly along $\mathcal{C}^1$. Therefore, we may globally define $G/f^1$ which is a section of $\mathcal{O}(\bar{K}-\mathcal{C}^1)$. If $G/f^1$ vanishes identically the CY threefold is singular along $\mathcal{C}^1\times \mathcal{C}^2$. Thus, we need that the line bundle $\mathcal{O}(\bar{K}-\mathcal{C}^1)$ has at least one non-trivial global section. This section is either constant, i.e. $[\bar{K}]=[\mathcal{C}^1]$, or it has a vanishing locus which is another effective curve $\tilde{\mathcal{C}}\sim \bar{K}-\mathcal{C}^1$. In the latter case, since $\mathcal{B}_1$ is Fano, and thus $\bar{K}\cdot \mathcal{C}>0$ for all effective curves $\mathcal{C}$, we must require that $\bar{K}\cdot (\bar{K}-\mathcal{C}^1)>0$.  This is not met by all the del-Pezzo involutions classified in \cite{Blumenhagen:2008zz}. 

The class II of orientifolds comes from involutions that interchange two isomorphic ambient space del Pezzo surfaces $\mathcal{B}_1\longleftrightarrow \mathcal{B}_2\simeq \mathcal{B}_1\equiv \mathcal{B}$. The fixed surface has the topology of $\mathcal{B}$, and the normal bundle is isomorphic to $T\mathcal{B}$. For an O3/O7 orientifold, the anti-canonical section has to be anti-symmetric across the fixed surface. Therefore, the number of conifold singularities is given by
\begin{equation}
n_{cf}=\int_{\mathcal{B}}c_2(\mathcal{B})-2c_1(\mathcal{B})^2+(2c_1(\mathcal{B}))^2=\chi_{\mathcal{B}}+2d_{\mathcal{B}}\, .
\end{equation}
The D3 tadpole is thus given by $-Q^{D3}=\frac{1}{4}(2\chi_{\mathcal{B}}+2d_{\mathcal{B}})=6$. A list of all the non-favourable orientifolds and some of their relevant data can be found \href{https://www.desy.de/~westphal/orientifold_webpage/cicy_orientifolds.html}{here}.

\subsubsection{The Schoen manifold}
The Schoen manifold is represented by the remaining 15 configuration matrices, and has $(h^{1,1},h^{2,1})=(19,19)$. A favourable description is given by the anti-canonical hypersurfaces in $dP_9\times dP_9$ where $dP_9$ is a rational elliptic surface, i.e. $\mathbb{P}^2$ blown up at nine points. As $h^{1,1}(dP_9\times dP_9)=20$, there must exists two distinct ambient space divisor classes that become equivalent once pulled back to the CY threefold, namely \cite{Anderson:2017aux}
\begin{equation}
[\bar{K}^1]|_{CY}=[\bar{K}^2]|_{CY}\, .
\end{equation}
Up to this subtlety, the discussion of orientifolds of the Schoen manifold is analogous to what we discussed in the previous section. Holomorphic involutions of $dP_9$ have been analyzed in \cite{Donagi:2000fw}, and we now give a very brief account of them, following \cite{Ovrut:2002jk}. $dP_9$ is an elliptic fibration over $\mathbb{P}^1$ with generically twelve degenerate fibers. It has $h^{1,1}=10$ and a standard basis of divisor classes $(l,e_1,...,e_9)$ that satisfy  
\begin{equation}
l^2=1\, ,\quad l\,e_i=0\, ,\quad e_ie_j=-\delta_{ij}\, ,
\end{equation}
where $l$ is the proper transform of the hyperplane class of the base $\mathbb{P}^1$ and the $e_{i}$ are the exceptional divisors associated with the blown up points.
The class of the generic fiber is the anti-canonical class $\bar{K}=3l-\sum_{i=1}e_i$, and the zero section can be identified with the exceptional divisor $e_9$. The fibration can be given a Weierstrass representation as follows: Consider a degree $(6,3)$ hypersurface in the toric variety specified by the GLSM 
\begin{equation}
\begin{pmatrix}
1 & 1 & 2 & 3 & 0\\
0 & 0 & 1 & 1 & 1
\end{pmatrix}\, .
\end{equation}
The first two coordinates $(t_1,t_2)$ are the projective $\mathbb{P}^1$ coordinates and the latter three are projective $\mathbb{P}^2$ coordinates that we label $(x,y,z)$, which are also sections of $(\mathcal{O}_{\mathbb{P}^1}(2),\mathcal{O}_{\mathbb{P}^1}(3),\mathcal{O}_{\mathbb{P}^1})$. $dP_9$ is the vanishing locus of
\begin{equation}
P\equiv -y^2z+x^3+g_2[t_1,t_2]xz^2+g_3[t_1,t_2]z^3=0\, ,
\end{equation}
with $g_2$ and $g_3$ generic degree four respectively six polynomials in the $\mathbb{P}^1$ coordinates.

The first basic involution is given by
\begin{equation}
(-1)_B:\, (t_1:t_2:x:y:z)\mapsto (t_1:t_2:x:-y:z)\, ,
\end{equation}
which acts trivially on the base and can be shown to reflect points in each fiber over the origin. Its fixed point divisor $D^o$ is the disjoint union of the zero section $e_9$ given by $x=z=0$ with the divisor $D_y$ defined by the transverse intersection of $y=0$ with the hypersurface. This intersects the generic fiber at four points, so $\bar{K}\cdot D^0=4$. Moreover, $e_9^2=-1$ and $D_y^2=9$, and there are no isolated fixed points.

The second basic involution is given by
\begin{equation}
\alpha_B:\, (t_1:t_2:x:y:z)\mapsto (t_1:-t_2:x:y:z)\, ,
\end{equation}
and the fixed point set is $\{t_1=0\}\cup \{t_2=y=0\}\cup \{t_2=x=z=0\}$ which is the disjoint union of the fiber over the point $t_1=0$ in the base and four isolated fixed points on the fiber over $t_2=0$. The former is in the anti-canonical class so has self-intersection zero.

Two further involutions are obtained by composing the above with a translation along the fiber by a section $\xi$, denoted $t_{\xi}$. $t_{\xi}\circ (-1)_B$ is an involution for any section $\xi$, while $t_{\xi}\circ \alpha_B$ is only an involution if $\alpha_B(\xi)=(-1)_B(\xi)$. As shown in section 4 of \cite{Donagi:2000fw} such sections can be found at particular loci in the moduli space of $dP_9$. For one such choice, the action on cohomology of $(-1)_B,\alpha_B$ and $t_{\xi}$ has been fully worked out (see table 1 of \cite{Donagi:2000fw}). We have that $t_{\xi}\circ \alpha_B$ has no fixed divisor but four isolated fixed points on the fiber $t_1=0$. Finally, $t_{\xi}\circ (-1)_B$ has a fixed divisor that intersects the generic fiber at four points, and whose class can be identified with $2(3l-\sum_{i=2}^8e_i)$ by noting the following. According to table 1 of \cite{Donagi:2000fw} $t_{\xi}\circ (-1)_B$ exchanges $e_1\leftrightarrow e_9$. As these are exceptional divisors, the involution in fact exchanges the two exceptional curves which do not intersect the fixed locus because $e_1e_9=0$. Thus, we can blow down both exceptional divisors and obtain an involution of $dP_7$ with no isolated fixed points. By comparing the action on cohomology with the results of \cite{Blumenhagen:2008zz} one finds that the involution one ends up after the blow-down indeed has one fixed divisor in $2(3l-\sum_{i=1}^7e_i)$ without isolated fixed points. We summarize these results in table \ref{tab:dP9involutions}.
\begin{table}
	\centering
	\begin{tabular}{c | c | c | c | c }
		involution & fixed divisor $D^o$ & $D^o\cdot c_1$ & $N_{fp}$ & $h^{1,1}_-$ \\ \hline
		$(dP_9,(-1)_B)$ & $e_9\cup D_y$ & $4$ & $0$ & $6$\\ 
		$(dP_9,(-\alpha)_B)$ & $3l-\sum_{i=1}^{9}e_i$ & $0$ & $4$ & $4$\\
		$(dP_9,t_{\xi}\circ(-1)_B)$ & $2(3l-\sum_{i=2}^{8}e_i)$ & $4$ & $0$ & $8$\\ 
		$(dP_9,t_{\xi}\circ(-\alpha)_B)$ & - & - & $4$ & $4$\\
	\end{tabular}
	\caption{Some $\mathbb{Z}_2$ involutions of $dP_9$ surfaces, the fixed divisor $D_1^o$, its intersection with the anti-canonical divisor $D_1^o\cdot c_1$, the number of isolated fixed points $N_{fp}$, and the values of $h^{1,1}_-$.}
	\label{tab:dP9involutions}
\end{table}
The resulting orientifolds, are grouped with the other \href{https://www.desy.de/~westphal/orientifold_webpage/cicy_orientifolds.html}{orientifolds of non-favourable CICYs}.
\section{Application: Ultra light throat axions}\label{sec::throataxions}
The CICYs are famously known to form a web of geometric phases connected to each other via conifold transitions \cite{Candelas:1989ug}. At the level of the configuration matrices, these are described by \textit{determinantal splittings}
\begin{equation}\label{eq:det-splitting}
\begin{bmatrix}
M & a_1 & \cdots & a_{n+1}\\
0 & 1   & \cdots & 1
\end{bmatrix}\longrightarrow 
\begin{bmatrix}
M & \sum_{i=1}^{n+1} a_i
\end{bmatrix}
\end{equation}
The manifolds described by the matrices on both sides are inequivalent if and only if their Euler numbers are different. If they are, the splitting is called \textit{effective}. Following through such a transition, $h^{1,1}$ decreases while $h^{2,1}$ increases. At the level of type II CY compactifications the low energy physics near the conifold transition loci was understood by Strominger \cite{Strominger:1995cz} and Greene, Morrison, and Strominger \cite{Greene:1995hu}: The deformation branch can be thought of as the Coulomb branch of the bulk $U(1)^{h^{2,1}}$ gauge theory, and the resolution branch corresponds to a partial Higgsing.

At the level of $O7$ orientifolded $\mathcal{N}=1$ flux compactifications there are arise new interesting phenomena: Generic three-form fluxes stabilize all complex structure moduli \cite{Gukov:1999ya,Giddings:2001yu} and more so, generically stabilize them exponentially close to conifold points in moduli space \cite{Giddings:2001yu,Denef:2004ze}. Backreaction of fluxes replaces the conifold regions by warped throats with exponential red-shifting \cite{Klebanov:2000hb,Giddings:2001yu}. 

Moreover, momentarily disregarding the truncation of the spectrum due to orientifolding, the change in $h^{1,1}$ across the $\mathcal{N}=2$ transition locus counts the number of exponentially light complex axions (\textit{thraxions}) whose field excursion weakly twists the various throats against each other \cite{Hebecker:2015tzo,Hebecker:2018yxs}. These axions can be identified with the massless axions one obtains on the resolved side of the transition by integrating $C_2$ and $B_2$ over the independent resolution two-cycles. On the deformed side they receive an exponentially small superpotential of order the deformation parameter. 

However, there are two necessary conditions that have to be satisfied in order for the thraxions to actually be part of the light spectrum. First, at the $\mathcal{N}=2$ level, it must be possible to cross the transition locus in a way that preserves a geometric $\mathbb{Z}_2$ symmetry, namely the one we would like to use to define an O3/O7 orientifold projection. In other words, across the locus $h^{1,1}_+$ must increase toward the resolved side. Second, the set of extra light axions one obtains on the resolved side of the transition must not be fully projected out by the O7 orientifold projection, i.e. $h^{1,1}_-$ must also increase by at least one toward the resolved side. Thus, two \textit{independent} resolution cycles must shrink to a set of conifold singularities, and the geometric $\mathbb{Z}_2$ involution associated with the O3/O7 orientifold must interchange them. 

It is easy to find explicit conifold transitions between orientifolds of anti-canonical surfaces in products of del Pezzo surfaces that satisfy all properties. As recalled in appendix \ref{app:CY-in-dP} a blow up/down of exceptional divisors in either of the two del Pezzo factors generically descends to a conifold transition in the CY threefold. Thus, we are instructed to inspect the web of $\mathbb{Z}_2$-compatible blow up/down transitions which has been listed in Figure 10 of \cite{Blumenhagen:2008zz}, and find transitions across which $h^{1,1}_-$ jumps at least by one. One caveat that we need to pay attention to is the following. If two $\mathbb{Z}_2$ symmetric del Pezzo surfaces that are connected by a blow up/down transition have different number of fixed divisors and/or isolated fixed points, the number of O-planes will jump across the transition locus. Thus, in the vicinity of the transition locus, some of the stacks are on the brink of merging or collapsing. While it would be interesting to analyze the spectrum of light degrees of freedom that appears when a CY orientifold is stabilized in such a region of moduli space, for now we would like to avoid this complication. Thus, we will restrict to blow-up/down transitions across which both the number of isolated fixed points and connected fixed divisors is invariant. We then expect that the seven-branes and O3 planes do not play an important role for the discussion. We list the blow up/down maps in table \ref{tab:delPezzo_transitions1}
\begin{table}
	\centering
	\begin{tabular}{ c | c | c}
		\text{del-Pezzo transitions} & \# fixed divisors & \# fixed points \\ \hline
		$(dP_4,\sigma_2)\longrightarrow(dP_2,\sigma_3)\longrightarrow (\mathbb{P}^2,\sigma)$& $1$ & $1$\\
		$(dP_5,\sigma_1)\longrightarrow(dP_3,\sigma_2)\longrightarrow (dP_1,\sigma_2)$& $1$ & $2$\\
		$(dP_5,\sigma_2)\longrightarrow(dP_3,\sigma_3)\longrightarrow (\mathbb{P}^1\times\mathbb{P}^1,\sigma_3)$& $1$ & $0$\\
		$(dP_6,\sigma_1)\longrightarrow(dP_4,\sigma_1)\longrightarrow (dP_2,\sigma_2)$& $1$ & $3$\\
		$(dP_7,\sigma_1)\longrightarrow(dP_5,\sigma_3)\longrightarrow (dP_3,\sigma_4)$& $0$ & $4$\\
		$(dP_7,\sigma_3)\longrightarrow(dP_5,\sigma_{dJ})$& $1$ & $0$\\
		$(dP_8,\sigma_2)\longrightarrow(dP_6,\sigma_2)$ & $1$ & $1$		
	\end{tabular}
	\caption{The $\mathbb{Z}_2$-invariant del-Pezzo transitions that lead to conifold singularities in the associated CY threefold that avoid the O-plane loci. Across each transition both $h^{1,1}_+$ and $h^{1,1}_-$ decrease by one. In the first five lines, two subsequent transitions are possible leading to jumps of $h^{1,1}_{\pm}$ by two.}
	\label{tab:delPezzo_transitions1}
\end{table}

Let us look in detail at the CY orientifold transitions obtained from the anti-canonical hypersurfaces in the chain of transitions 
\begin{align}
\begin{tabular}{c}
$(dP_5,\sigma_2)\times (\mathbb{P}^2,\id)$ \\ \hline
$(h^{1,1}_+,h^{1,1}_-)=(4,3)$\\
$(h^{2,1}_+,h^{2,1}_-)=(11,32)$
\end{tabular}\, \longrightarrow \,
\begin{tabular}{c}
$(dP_3,\sigma_3)\times (\mathbb{P}^2,\id)$ \\ \hline
$(h^{1,1}_+,h^{1,1}_-)=(3,2)$\\
$(h^{2,1}_+,h^{2,1}_-)=(19,40)$
\end{tabular} \, \longrightarrow \,  
\begin{tabular}{c}
$(\mathbb{P}^1\times \mathbb{P}^1,\sigma_3)\times (\mathbb{P}^2,\id)$ \\ \hline
$(h^{1,1}_+,h^{1,1}_-)=(2,1)$\\
$(h^{2,1}_+,h^{2,1}_-)=(27,48)$
\end{tabular}
\end{align}
In each of the three orientifolds the D3 tadpole is equal to 12, and there are no conifold singularities at generic points in moduli space. Across each of the two transitions there appear 9 conifold singularities (18 in the double cover) that are not located on top of O7 planes and that can subsequently be deformed in a way that is compatible with the orientifolding. Locally, this looks just like the $\mathcal{N}=2$ conifold transitions. Upon stabilizing the deformation modes near one or both of the conifold singularities, one obtains strongly warped throats with one respectively two independent complex axions with exponentially small superpotential as described in \cite{Hebecker:2018yxs}. 

For the orientifolds of favourable CICYs, one searches for involutions that swap at least one pair of rows, each containing only one's and zero's, with at most one common non-vanishing entry. If this is the case, the first row can be collapsed using \eqref{eq:det-splitting}, such that the second row that is obtained after the splitting still contains only one's and can thus be collapsed as well. There are $319,521$ orientifolds in our list that fulfill this requirement (see \href{https://www.desy.de/~westphal/orientifold_webpage/cicy_orientifolds.html}{\textit{thraxion\_candidates}}). These are $\sim 94\%$ of the cases with $h^{1,1}_-\neq 0$. In addition, analogously to what we have said for the conifold transitions in the non-favourable cases, one might want to further require that the O3 and O7 planes are not transformed by the conifold transition. Filtering out these, we obtain a list of 11,533 orientifold conifold transitions\footnote{This list is not complete because we have only considered cases where a suitably obvious set of swaps of rows and columns brings the configuration matrix obtained via the splitting to the form of one of the configuration matrices recorded in the CICY-list.} (\href{https://www.desy.de/~westphal/orientifold_webpage/cicy_orientifolds.html}{\textit{thraxion\_transitions}}) each encoded in an entry
\begin{equation}
\{\{i_{CICY},i_{o-fold},i_{row-pair}\},\{j_{CICY},j_{o-fold}\}\}\, ,
\end{equation}
where the first entry corresponds to the resolved side and contains the CICY number $i_{CICY}$, the position of the orientifold in our list $i_{o-fold}$, and an index $i_{row-pair}$ indicating that the $i_{row-pair}$~-th pair of rows interchanged by the involution is collapsed. The second entry specifies the deformed side CICY-orientifold. 1,279 of these are transitions between pairs of orientifolds of CICYs that are smooth away from the transition locus, as opposed to orientifolds of the larger set of smooth CYs that are reached from the CICYs via the resolution of frozen conifold singularities. Here is an example,
\begin{align}
\left[\begin{tabular}{c | c c c c}
$\mathbb{P}^{2}$ & $1$ & $1$ & $1$ & $0$  \\
$\mathbb{P}^{1}$ & $0$ & $0$ & $1$ & $1$ \\
$\mathbb{P}^{1}$ & $0$ & $1$ & $0$ & $1$ \\
$\mathbb{P}^{1}$ & $1$ & $0$ & $0$ & $1$ \\
$\mathbb{P}^{2}$ & $0$ & $0$ & $0$ & $3$ \\
\end{tabular}\right]&\longrightarrow 
\left[\begin{tabular}{c | c c }
$\mathbb{P}^{1}$ & $1$ & $1$   \\
$\mathbb{P}^{2}$ & $1$ & $2$    \\
$\mathbb{P}^{2}$ & $0$ & $3$ \\
\end{tabular}\right]
\, .
\end{align}
On the resolved side, the second and third row as well as the second and third column are interchanged by the involution, $\mathcal{I}=(1,0,0,0,0)$, and all parities are positive. The deformed side is reached by collapsing the two interchanged rows. No further rows and columns are interchanged, $\mathcal{I}=(0,1,0)$, and again all parities are positive. On both sides there are two smooth O7 divisors with $(\chi_1,d_1)=(36,0)$ and $(\chi_2,d_2)=(12,0)$ respectively, so the D3-tadpole is $12$. The Hodge numbers $(h^{1,1}_+,h^{1,1}_-,h^{2,1}_+,h^{2,1}_-)$ transform as
\begin{equation}
(4,1,20,39)\longrightarrow (3,0,28,47)\, ,
\end{equation}
and on the transition locus of the double-cover of the orientifold there are $|\Delta \chi_{CY}|/2=18$ conifold singularities with two homology relations among the shrinking three-spheres. These can be grouped into a pair of 9 conifolds each that is interchanged by the orientifold action.

We leave the phenomenology of this interesting class of axion-models for future work. Here, we note only that the examples with two axions we have considered feature a diagonal kinetic matrix. Then, at the classical level, the two axion-sectors are sequestered from one-another in that their scalar potentials simply add up.

\vspace{0.3cm}

\noindent\textbf{Acknowledgements:} We are grateful for useful discussions with Andreas Braun, Mehmet Demirtas, Naomi Gendler, Mariana Graña, Arthur Hebecker, Manki Kim, Liam McAllister, Nicole Righi, Fabian Rühle, Raffaele Savelli, Cumrun Vafa, and Timo Weigand. We are particularly indebted to Andreas Braun for collaboration in the early stages of the project and for many helpful discussions. JM and AW would like to thank the KITP Santa Barbara for its warm hospitality while this work was finished. The work of JM was supported in part by the Simons Foundation Origins of the  Universe  Initiative and in part by the ERC Consolidator Grant STRINGFLATION under the HORIZON 2020 grant agreement no. 647995. FC and AW are supported by the ERC Consolidator Grant STRINGFLATION under the HORIZON 2020 grant agreement no. 647995. The work of AW was also partially supported by the Deutsche Forschungsgemeinschaft under Germany’s Excellence Strategy - EXC 2121 “Quantum Universe” - 390833306. Finally, this work was supported in part by the National Science Foundation under Grant No. PHY-1748958.

\appendix
\section{Basic properties of del Pezzo surfaces}\label{app:dPproperties}

In this appendix we recall a quick collection of basic facts about algebraic geometrical properties of del Pezzo surfaces. See for example \cite{Hubsch:1992nu} for a more detailed treatment.

By definition a Fano variety $X$ over an algebraically closed field $\mathbb{F}$ is a complete irreducible algebraic variety such that its anticanonical sheaf $\bar{K}_X$ is ample. In the following we will only focus on Gorenstein Fano varieties, so the anticanonical bundle exists and by definition it has to be ample. The Nakai–Moishezon criterion implies that the ampleness of the anticanonical bundle is equivalent to the requirement that for every effective curve $\mathcal{C}$, the intersection with the anti-canonical divisor $\bar{K}$ is positive, i.e. $\bar{K} \cdot \mathcal{C}>0$. A del Pezzo surface is a smooth projective Fano variety of Krull dimension $2$.

From now on, we will only consider del Pezzo surfaces over the field $\mathbb{C}$, and Krull dimension coincides with ordinary dimension in differential geometry. Complex del Pezzo surfaces can be classified as follows: we can either take $\mathbb{P}^2$ and blow up $k=0,\cdots 8$ points, or we can take $\mathbb{P}^1\times\mathbb{P}^1$ and blow up $k=0,\cdots 7$ generic points. We will call the first class of surfaces $dP_k$ and the latter $F_k$. It is well known that $F_k\simeq B_{k+1}$, so in particular it is enough to consider just $\mathbb{P}^1\times \mathbb{P}^1$ and $dP_k$. Note that trivially $dP_0\simeq \mathbb{P}^2$. In the following we will denote with $\mathcal{B}$ any of those 10 del Pezzo surfaces.

Let us discuss the topological properties of $\mathcal{B}$, and let us start with $dP_k$.
We denote by $l$ the proper transform of the hyperplane class of the $\mathbb{P}^2$, and $e_i$ the exceptional divisors, associated to the blown-up points. The intersection numbers read
\begin{equation}
l\cdot l=1, \quad l\cdot e_i=0, \quad e_i\cdot e_j=-\delta_{ij}, \quad i,j=1,\cdots, k
\end{equation}
The anti-canonical class of the del Pezzo surfaces is given by
\begin{equation}
	\bar{K}=c_1(dP_k)=3l-\sum_{i=1}^{k}e_i \, ,
\end{equation}
and its Euler characteristic is $3+k$. Its \textit{degree} is 
\begin{equation}
d=\int_{dP_k}c_1(dP_k)^2=9-k\, ,
\end{equation}
and measures the number of points where two generic anti-canonical divisors intersect.
For $k=0,1,2,3$ the $dP_k$ are toric varieties, while for higher $k$ this is no longer true. Let us briefly move to $\mathbb{P}^1\times \mathbb{P}^1$ now. Being the direct product of two projective planes, it has two hyperplane classes $H$ and $H'$, first Chern class $c_1=2H+2H'$, degree $d=8$, and it also admit a toric description. 

\begin{figure}
	\centering
	\begin{tabular}{c c c}
	\includegraphics[keepaspectratio,width=5cm,clip,trim={3cm 12cm 5cm 3cm}]{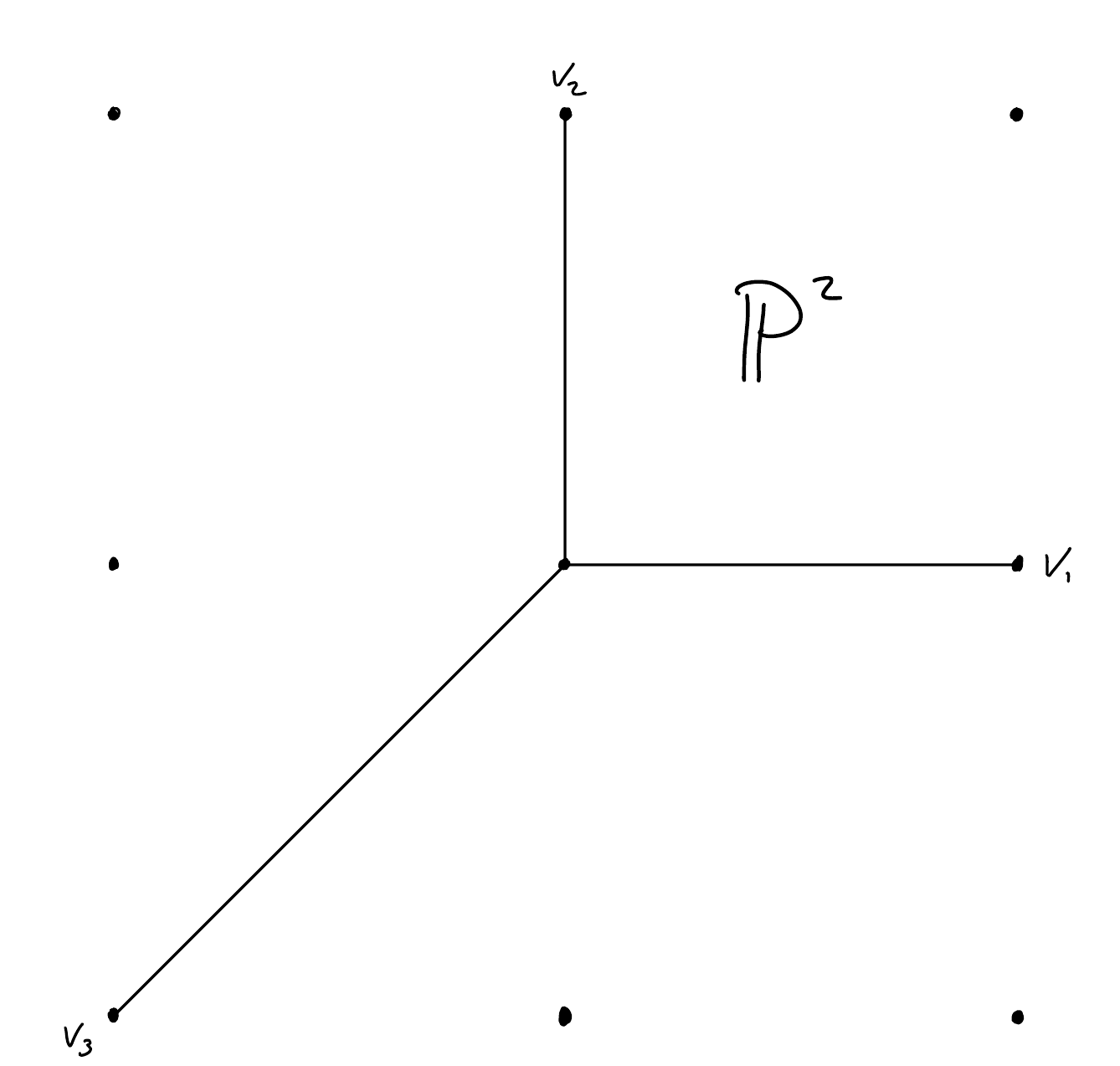}&  
	\includegraphics[keepaspectratio,width=5cm,clip,trim={3cm 12cm 5cm 3cm}]{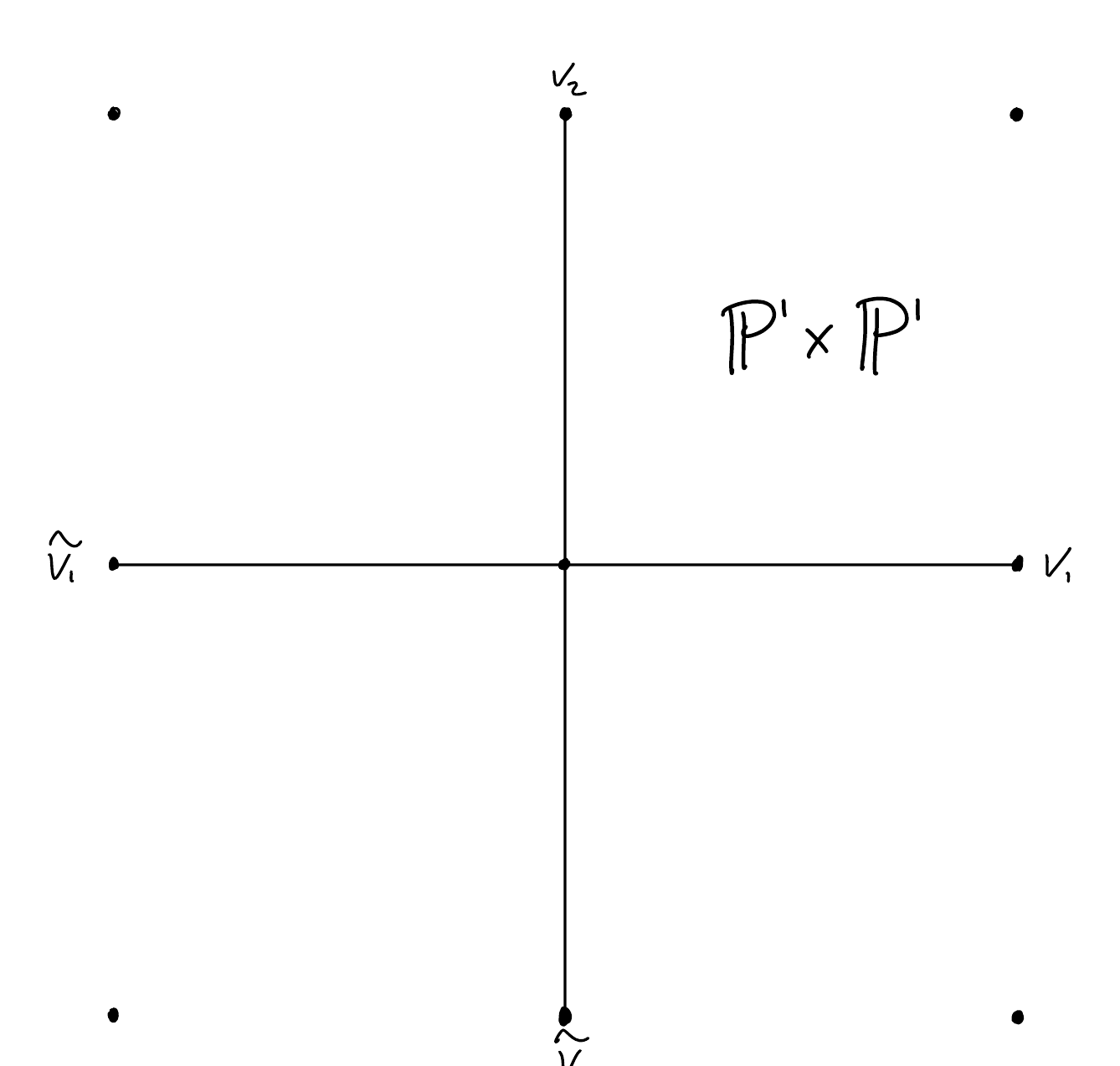}&
	\includegraphics[keepaspectratio,width=5cm,clip,trim={3cm 12cm 5cm 3cm}]{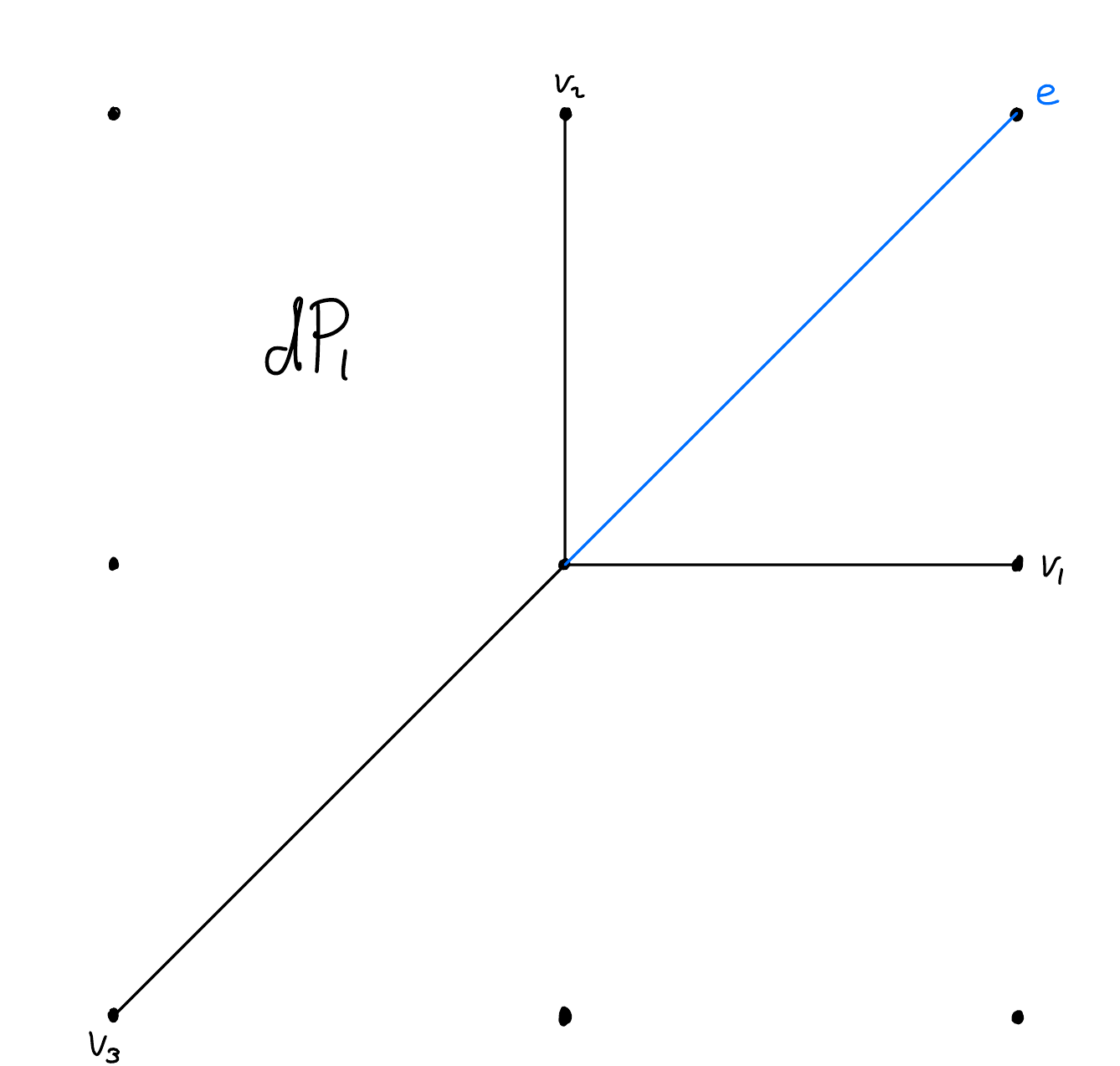} \\
	\includegraphics[keepaspectratio,width=5cm,clip,trim={3cm 12cm 5cm 3cm}]{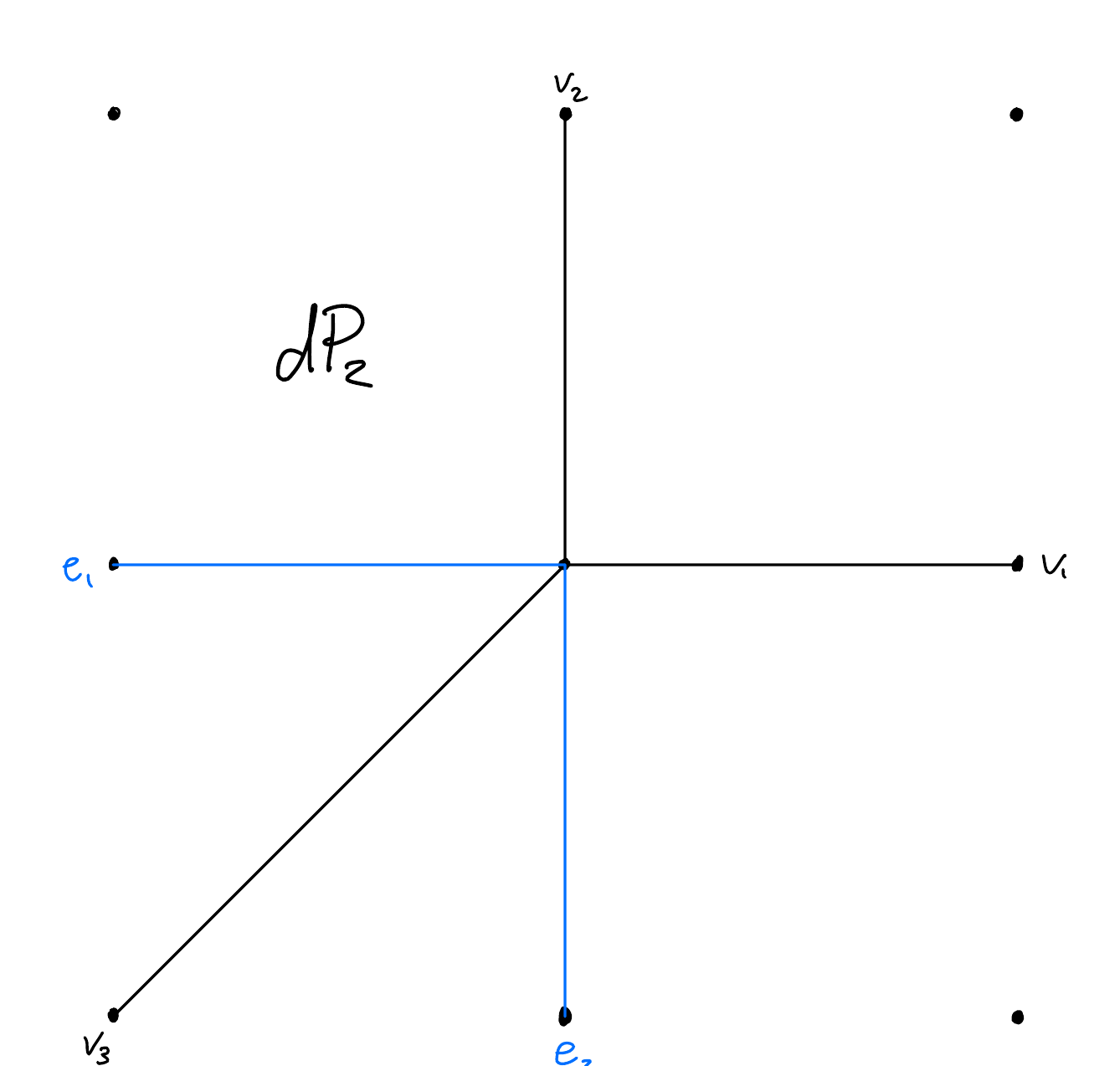}&
	\includegraphics[keepaspectratio,width=5cm,clip,trim={3cm 12cm 5cm 3cm}]{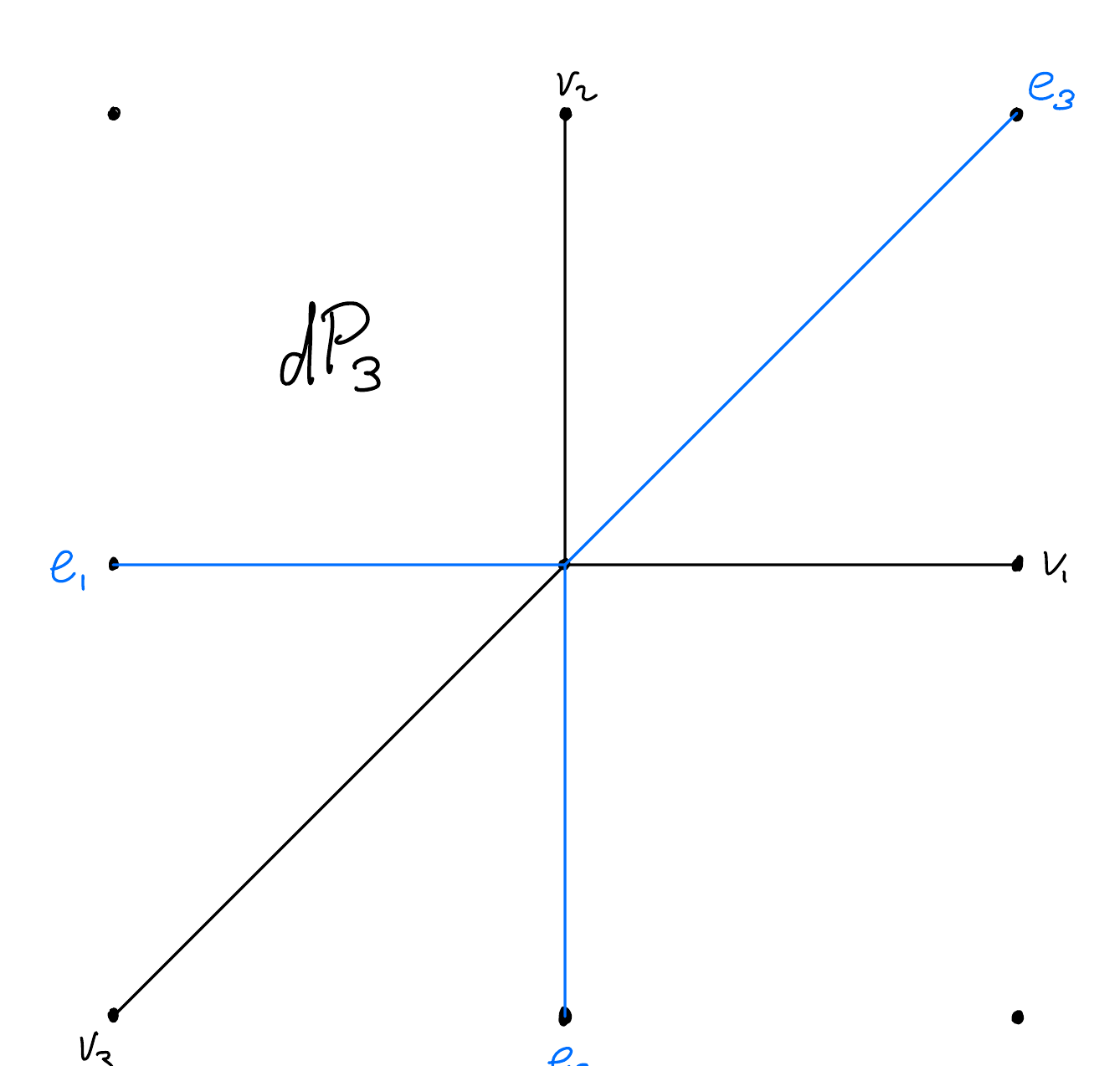}
	\end{tabular}
	\caption{The fans of the toric del Pezzo surfaces. The blue edges are associated with toric coordinates $e_i$ s.t. the locus $e_i=0$ corresponds to the exceptional divisor $E_i$.}
	\label{fig:fan_toricdelPezzos}
\end{figure}

The fans for the toric del Pezzos $dP_0,...,dP_3$ and $\mathbb{P}^1\times \mathbb{P}^1$ are given in figure \ref{fig:fan_toricdelPezzos},
and the associated GLSM charge matrices are
\begin{align}
Q_{\mathbb{P}^2}&=\begin{pmatrix}
1 \\ 
1 \\
1 
\end{pmatrix}\, ,\quad
\quad 
Q_{\mathbb{P}^1\times \mathbb{P}^1}=\begin{pmatrix}
1 & 0 \\ 
1 & 0 \\
0 & 1 \\
0 & 1 
\end{pmatrix}\, , \quad
Q_{dP_1}=\begin{pmatrix}
1 & 0 \\ 
1 & 1 \\
1 & 1 \\
0 & -1
\end{pmatrix}\, , \nonumber \\
Q_{dP_2}&=\begin{pmatrix}
1 & 0 & 1\\ 
1 & 1 & 0\\
1 & 1 & 1\\
0 & -1 & 0 \\
0 & 0 & -1
\end{pmatrix}\, , \quad 
Q_{dP_3}=\begin{pmatrix}
1 & 0 & 1 & 1\\ 
1 & 1 & 0 & 1\\
1 & 1 & 1 & 0\\
0 & -1 & 0 & 0\\
0 & 0 & -1 & 0\\
0 & 0 & 0 & -1
\end{pmatrix}\, .
\end{align}
The automorphism groups $\text{Aut}(\mathcal{B})$ are easily determined from the automorphism group of $\mathbb{P}^2$ which is $PGL(3,\mathbb{C})$. Their image under the blow-down map $dP_k\longrightarrow \mathbb{P}^2$ is simply the subgroup of  $PGL(3,\mathbb{C})$ that leaves the set of $k$ marked points (the ones that are blown up) invariant. 
Their dimensions are
\begin{equation}
\text{dim Aut}(\mathcal{B})=\max\{2d-10,0\}\, ,
\end{equation}
which holds also for $\mathbb{P}^1\times \mathbb{P}^1$.
We would also like to know the number of complex structure moduli of a del Pezzo surface $h_{cs}(\mathcal{B})=\text{dim}\,H^1(\mathcal{B},T\mathcal{B})$. As each of the $k$ blown up points is labeled by two complex coordinates we have $h_{cs}(\mathcal{B}) = \max\{2k-\text{dim Aut}(\mathbb{P}^2),0\}$, i.e.
\begin{equation}\label{eq:complexstructuresofdelPezzo}
h_{cs}(\mathcal{B})=
\max\{10-2d,0\}\, ,
\end{equation}
which again holds for $\mathbb{P}^1\times \mathbb{P}^1$ as well. We may also write
\begin{equation}
h_{cs}(\mathcal{B})-\text{dim Aut }(\mathcal{B})=10-2d\,  .
\end{equation}
A further important quantity is the dimension of the space of global sections of the anti-canonical line bundle on a del Pezzo surface $\mathcal{B}$. The Grothendieck-Riemann-Roch theorem says that
\begin{align}
\text{dim}\,H^{0}(\mathcal{B},\bar{K})&-\text{dim}\,H^{1}(\mathcal{B},\bar{K})+\text{dim}\,H^{2}(\mathcal{B}, \bar{K})\nonumber\\
&=\int ch(\bar{K})Td(\mathcal{B})=d+1\,
\label{GRR}
\end{align}
where $ch(\bar{K})$ is the Chern class of the anticanonical bundle, and $Td(\mathcal{B})$ the Todd class of (the tangent bundle on) $\mathcal{B}$.
The Kodaira-vanishing theorem implies the vanishing of $\text{dim}\, H^i(\mathcal{B},K\otimes \mathcal{L})$ for all $i>0$ if the line bundle $\mathcal{L}$ is positive. Choosing $\mathcal{L}=2\bar{K}$ which is positive for Fano spaces, we deduce from (\ref{GRR}) that in fact
\begin{equation}\label{eq:global_sections}
\text{dim}\,H^{0}(\mathcal{B},\bar{K})=d+1\, .
\end{equation}

Finally, we recall that all the del Pezzo surfaces with degree $d\neq 1$ can be written as a complete intersection of hypersurfaces in a product of projective spaces. This is why they appear in the CICY list. One of many ways to represent them is as follows,
\begin{align}
dP_1&=\left[\begin{tabular}{c | c}
$\mathbb{P}^1$ & $1$ \\
$\mathbb{P}^2$ & $1$ 
\end{tabular}\right]\, ,\quad 
dP_2=\left[\begin{tabular}{c | c c}
$\mathbb{P}^1$ & $1$ & $0$ \\
$\mathbb{P}^1$ & $0$ & $1$ \\
$\mathbb{P}^2$ & $1$ & $1$
\end{tabular}\right]\, ,\quad 
dP_3=\left[\begin{tabular}{c | c c}
$\mathbb{P}^2$ & $1$ & $1$ \\
$\mathbb{P}^2$ & $1$ & $1$
\end{tabular}\right]\, , \nonumber \\
dP_4&=\left[\begin{tabular}{c | c}
$\mathbb{P}^1$ & $1$ \\
$\mathbb{P}^2$ & $2$ 
\end{tabular}\right]\, ,\quad 
dP_5=\left[\begin{tabular}{c | c c}
$\mathbb{P}^4$ & $2$ & $2$
\end{tabular}\right]\, ,\quad 
dP_6=\left[\begin{tabular}{c | c}
$\mathbb{P}^3$ & $3$ 
\end{tabular}\right]\, , \quad
dP_7=\left[\begin{tabular}{c | c}
$\mathbb{P}^1$ & $2$ \\
$\mathbb{P}^2$ & $2$ 
\end{tabular}\right]\, .
\end{align}

\section{The anticanonical hypersurface in a product of two del Pezzos}\label{app:CY-in-dP}

In the main text, we are interested in CY threefolds that arise as anti-canonical divisors in $\mathcal{B}_1\times \mathcal{B}_2$ with $\mathcal{B}_{1,2}$ del Pezzo, and in determining their number of K\"ahler and complex structure moduli. In this appendix we fill in some details. Let $d_1$ (resp. $d_2$) be the degree of $\mathcal{B}_1$ (resp $\mathcal{B}_2$).

For the K\"ahler moduli, we will use the following theorem of Kollár \cite{Borcea:1991aaa}:
\begin{Theorem}
	(Koll\'ar) Let $X$ be a smooth Fano variety with $\rm{dim}(X) \geq 4$. Let $Y \subset X$ be a smooth divisor in the class $\bar{K}_X$. Let $\overline{NE}(Y)$ (resp. $\overline{NE}(X)$) be the Kleiman-Mori cone of $Y$ (resp. $X$). Then the natural inclusion
	\begin{equation}
	i_*: \overline{NE}(Y) \to \overline{NE}(X)
	\end{equation}
	is an isomorphism.
\end{Theorem}
This theorem implies that in our case under study, all the K\"ahler moduli of the CY threefold simply descend from the ambient space ones, so
\begin{equation}
h^{1,1}=20-d_1-d_2\, .
\end{equation}

The number of complex structure moduli can be determined from the Euler characteristic by subtraction. Using the adjunction formula one determines $\chi_{CY}=-2d_1d_2$ and therefore
\begin{equation}
h^{2,1}=20-d_1-d_2+d_1d_2\, .
\end{equation}

It is useful to rewrite this as follows
\begin{align}
h^{2,1}=&\text{dim}\,H^{0}(\mathcal{B}_1,\bar{K}) \cdot \text{dim}\,H^{0}(\mathcal{B}_2,\bar{K})+\nonumber\\
&-1-\text{dim Aut}(\mathcal{B}_1)-\text{dim Aut}(\mathcal{B}_2)\nonumber+\\
&+h_{cs}(\mathcal{B}_1)+h_{cs}(\mathcal{B}_2)\, .
\label{eq:h21dpdp}
\end{align}
Equation \ref{eq:h21dpdp} implies that we can think of all of the $h^{2,1}$ complex structure moduli as coefficients appearing in a general global section of the anti-canonical line bundle of $\mathcal{B}_1\times \mathcal{B}_2$, subtracting one for the overall scaling and the dimension of the ambient space automorphism group, and finally adding the complex structure moduli inherited from the ambient space. Clearly, if we want to compute $h^{2,1}_-$ for a given $\mathbb{Z}_2$ involution acting only on the first factor $\mathcal{B}_1$, we replace $H^{0}(\mathcal{B}_1,\bar{K})$ by the vector space of $\mathbb{Z}_2$-even global sections $H^{0}_+(\mathcal{B}_1,\bar{K})$, replace $\text{Aut}(\mathcal{B}_1)$ by the subgroup of the automorphism group that commutes with the $\mathbb{Z}_2$ action, denoted $\text{Aut}_+(\mathcal{B}_1)$, and replace $h_{cs}(\mathcal{B}_1)$ by $h_{cs}^+(\mathcal{B}_1)$. This way of computing $h^{2,1}_-$, rather than employing the Lefschetz fixed point formula \eqref{eq:LefschetzFixedPoint}, is employed as a cross-check.

Furthermore, it is important to note that upon blowing down an exceptional divisor of the first ambient space factor $\mathcal{B}_1$, one obtains $d_2=\int_{\mathcal{B}_2}c_1(\mathcal{B}_2)^2$ conifold singularities in the CY. This is seen most easily as follows: Every exceptional divisor of $dP_k$ intersects the anti-canonical divisor at one point, i.e. $c_1(dP_k)\cdot E_i=1$. Thus, upon blowing down an exceptional divisor to a point, thus creating $dP_{k-1}$, the anti-canonical divisor will contain this point. This is why dim $H^0(\mathcal{B},\bar{K})$ increases with the degree. This point is given by the intersection of two divisors $D_1$ and $D_2$ in $dP_{k-1}$ that are represented as the vanishing loci of global sections $p_{1,2}$ of the respective divisor line bundles. Because this point is contained in the anti-canonical divisor we started with, the anti-canonical divisor is represented by a global section $f$ of the anti-canonical line bundle that can be expanded as\footnote{Of course, also before the blow-down the generic section can be expanded as in eq. \eqref{eq:genericsection}, but the intersection $D_1\cdot D_2$ is empty in $dP_k$, so there are no conifold singularities.}
\begin{equation}\label{eq:genericsection}
f=p_1\cdot g_1+p_2\cdot g_2\, ,
\end{equation}
where $g_1$ and $g_{2}$ are global sections of the anti-canonical line bundle of $\mathcal{B}_2$ with coefficients that are global sections of a line bundle $\mathcal{L}_1$ respectively $\mathcal{L}_2$ on $dP_{k-1}$. Generically, the common vanishing locus $p_1=p_2=g_1=g_2=0$ produces a conifold singularity in the CY because all four sections will vanish linearly. The number of such points is given by the generic number of intersections between two anti-canonical divisors in $\mathcal{B}_2$, i.e. by the degree $d_2$ of $\mathcal{B}_2$. Of course, once the exceptional divisor is blown down, we can deform the singularity by dialing the $d_2-1$ newly acquired complex structure moduli. Thus, we have described a conifold transition in the CY from the resolved to the deformed side across which $h^{1,1}$ decreases by one, and $h^{2,1}$ increases by $d_2-1$. Since $dP_k$ is $\mathbb{P}^2$ blown up at generic points, no pair of them coincides. Therefore, we may blow down any number $n\leq k$ of exceptional divisors \textit{without} producing more severe singularities than conifolds in the threefold. Using this, one obtains conifold transitions across which $h^{1,1}$ decreases by $n$ and $h^{2,1}$ increases by $n(d_2-1)$.

\section{Involutions of del Pezzo surfaces}
The involutions of del Pezzo surfaces have been classified in \cite{Blumenhagen:2008zz}, and are listed in table \ref{tab:dPinvolutions}
\begin{table}
	\centering
	\begin{tabular}{c | c | c | c | c | c}
		involution & fixed divisor $D^o$ & $D^o\cdot c_1$ & $N_{fp}$ & $h^{1,1}_-$ & action on $H_2$ \\ \hline
		$(\mathbb{P}^2,\sigma)$ & $l$ & $3$ & $1$ & $0$ & $\id_1$\\ \hline
		$(\mathbb{P^1}\times \mathbb{P^1},\sigma_1)$ & $(H)\cup (H')$ & $4$ & $0$ & $0$ & $\id_2$\\ 
		$(\mathbb{P^1}\times \mathbb{P^1},\sigma_2)$ & - & - & $4$ & $0$ & $\id_2$\\
		$(\mathbb{P^1}\times \mathbb{P^1},\sigma_3)$ & $H+H'$ & $4$ & $0$ & $1$ & $F$\\ \hline
		$(dP_1,\sigma_1)$ & $(l)\cup (e_1)$ & $4$ & $0$ & $0$ & $\id_2$\\  
		$(dP_1,\sigma_2)$ & $l-e_1$ & $2$ & $2$ & $0$ & $\id_2$\\ \hline
		$(dP_2,\sigma_1)$ & $(l-e_1)\cup (e_2)$ & $3$ & $1$  & $0$ & $\id_3$\\ 
		$(dP_2,\sigma_2)$ & $l-e_1-e_2$ & $1$ & $3$  & $0$ & $\id_3$\\
		$(dP_2,\sigma_3)$ & $l$ & $3$ & $1$  & $1$ & $\id_1\oplus F$\\ \hline
		$(dP_3,\sigma_1)$ & $(l-e_1-e_2)\cup (e_3)$ & $2$ & $2$ & $0$ & $\id_4$\\ 
		$(dP_3,\sigma_2)$ & $l-e_3$ & $2$ & $2$ & $1$ & $\id_2\oplus F$\\ 
		$(dP_3,\sigma_3)$ & $2l-e_1-e_2$ & $4$ & $0$ & $2$ & $I_{dP_3}^{(2)}$\\ 
		$(dP_3,\sigma_4)$ & - & - & $4$ & $1$ & $I_{dP_3}^{(3)}$\\ \hline
		$(dP_4,\sigma_1)$ & $l-e_1-e_2$ & $1$ & $3$ & $1$ & $\id_3\oplus F$\\ 
		$(dP_4,\sigma_2)$ & $l$ & $3$ & $1$ & $2$ & $\id_1\oplus F\oplus F$\\ \hline
		$(dP_5,\sigma_1)$ & $l-e_1$ &  $2$ & $2$ & $2$ & $\id_2\oplus F \oplus F$\\ 
		$(dP_5,\sigma_2)$ & $2l-e_1-e_2$ & $4$ & $0$ & $3$ & $I_{dP_3}^{(2)}\oplus F$\\
		$(dP_5,\sigma_3)$ & - & - & $4$ & $2$ & $I_{dP_3}^{(3)}\oplus F$\\ 
		$(dP_5,\sigma_{\text{dJ}})$ & $3l-\sum_{i=1}^{5}e_i$ & $4$ & $0$ & $4$ & $I_{dP_5}^{(5)}$\\ \hline
		$(dP_6,\sigma_1)$ & $l-e_1-e_2$ & $1$ & $3$ & $2$ & $\id_3\oplus F \oplus F$\\ 
		$(dP_6,\sigma_2)$ & $3l-\sum_{i=1}^{6}e_i$ & $3$  & $1$ & $4$ & $I_{dP_5}^{(5)}\oplus \id_1$\\ \hline
		$(dP_7,\sigma_1)$ & - & -  & $4$ & $3$ & $I_{dP_3}^{(3)}\oplus F \oplus F$\\
		$(dP_7,\sigma_2)$ & $3l-\sum_{i=1}^{7}e_i$ & $2$  & $2$ & $4$ & $I_{dP_5}^{(5)}\oplus \id_2$\\ 
		$(dP_7,\sigma_3)$ & $3l-\sum_{i=1}^{5}e_i$ & $4$ & $0$ & $5$ & $I_{dP_5}^{(5)} \oplus F$\\ 
		$(dP_7,\sigma_G)$ & $6l-2\sum_{i=1}^{7}e_i$ & $4$ & $0$  & $7$ & $I_{dP_7}^{(9)}$\\ \hline
		$(dP_8,\sigma_1)$ & $3l-\sum_{i=1}^{8}e_i$ & $1$ & $3$ & $4$ & $I_{dP_5}^{(5)}\oplus \id_3$\\ 
		$(dP_8,\sigma_2)$ & $3l-\sum_{i=1}^{6}e_i$ & $3$ & $1$ & $5$ & $I_{dP_5}^{(5)} \oplus \id_1 \oplus F$\\ 
		$(dP_8,\sigma_B)$ & $9l-3\sum_{i=1}^{8}e_i$ & $3$ & $1$ & $8$ & $I_{dP_8}^{(9)}$\\ \hline
	\end{tabular}
	\caption{Table adopted from table 6 of \cite{Blumenhagen:2008zz}: The possible $\mathbb{Z}_2$ involutions of del Pezzo surfaces, the fixed divisor $D_1^o$, its intersection with the anti-canonical divisor $D_1^o\cdot c_1$, the number of isolated fixed points $N_{fp}$, the values of $h^{1,1}_-$, and actions on the divisor classes $(l,e_1,...,e_n)$ respectively $(H,H')$. $F$ denotes a \textit{flip} $e_i\longleftrightarrow e_j$, respectively $H\longleftrightarrow H'$.}
	\label{tab:dPinvolutions}
\end{table}

Let us explain in detail the possible $\mathbb{Z}_2$ involutions of the \textit{toric} del Pezzo surfaces, i.e. $\mathbb{P}\simeq dP_0,\mathbb{P}^1\times \mathbb{P}^1,dP_1,dP_2$ and $dP_3$. 
We associate a toric coordinate $x_\rho$ to each generator of a one-dimensional cone $\rho\in \Sigma(1)$ of the fan. 
It is useful to first consider the case of $dP_3$. There are four scaling relations among the six toric coordinates $\{x_{v_1},x_{v_2},x_{v_3},x_{e_1},x_{e_2},x_{e_3}\}$,
\begin{align}\label{eq:dP3_scaling}
(x_{v_1},x_{v_2},x_{v_3})&\sim \lambda_0 (x_{v_1},x_{v_2},x_{v_3})\, ,\nonumber\\
(x_{v_2},x_{v_3},x_{e_1})&\sim  (\lambda_1x_{v_2},\lambda_1x_{v_3},\lambda_1^{-1}x_{e_1})\, ,\nonumber\\
(x_{v_1},x_{v_3},x_{e_2})&\sim  (\lambda_2x_{v_1},\lambda_2x_{v_3},\lambda_2^{-1}x_{e_2})\, ,\nonumber\\
(x_{v_1},x_{v_2},x_{e_3})&\sim  (\lambda_3x_{v_1},\lambda_3x_{v_2},\lambda_3^{-1}x_{e_3})\, ,
\end{align}
To each toric coordinate we associate a toric divisor $D_{\rho}$, $\{D_{v_1},D_{v_2},D_{v_3},D_{e_1},D_{e_2},D_{e_3}\}$, that satisfy homology relations
\begin{align}
[D_{v_1}]+[D_{e_2}]&=[D_{v_2}]+[D_{e_3}]\nonumber\\
[D_{v_1}]+[D_{e_3}]&=[D_{v_3}]+[D_{e_1}]\nonumber\\
[D_{v_2}]+[D_{e_3}]&=[D_{v_3}]+[D_{e_2}]\, .
\end{align}
Let us work in a basis of divisor classes
\begin{align}
[l]&\equiv[D_{v_1}]+[D_{e_2}]+[D_{e_3}]=[D_{v_2}]+[D_{e_1}]+[D_{e_3}]=[D_{v_3}]+[D_{e_1}]+[D_{e_2}]\nonumber\\
[e_1]&=[D_{e_1}]\, ,\quad [e_2]=[D_{e_2}]\, ,\quad [e_3]=[D_{e_3}]\, .
\end{align}
We expand the K\"ahler form as $J=t^0 [l]-\sum_{i=0}^3t^i [e_i]$. All divisors have positive volume if
\begin{equation}
t^0-t^i-t^j>0\, , \quad i,j=1,2,3\, ,\quad  i\neq j\, , \text{and}\quad t^i>0 \, , \quad i=1,2,3\, .
\end{equation}
For $dP_2$ we delete the last row of \eqref{eq:dP3_scaling} and set $D_{e_3}\rightarrow 0$, for $dP_1$ we delete the second and third rows, set $D_{e_{1,2}}\rightarrow 0$ and relabel $e_{3}\longrightarrow e_1$, and for $dP_0\simeq \mathbb{P}^2$ only the first remains and all exceptional divisors are shrunken. $\mathbb{P}^1\times \mathbb{P}^1$ is obtained by shrinking $v_3$ and $e_3$. A geometric action on the toric ambient space coordinates translates to a geometric action on the fan. $dP_2$ admits a $\mathbb{Z}_2$ symmetry that exchanges
\begin{equation}
x_{v_1}\longleftrightarrow x_{v_2}\, ,\quad x_{e_1}\longleftrightarrow x_{e_2}\, .
\end{equation}
The $\mathbb{Z}_2$ fixed point locus $\{x_{v_1}x_{e_2}-x_{v_2}x_{e_1}=0\}\cup \{x_{v_1}x_{e_2}+x_{v_2}x_{e_1}=x_3=0\}$ is the disjoint union of a $\mathbb{P}^1$ and an isolated fixed point. The action on our basis of divisors is
\begin{equation}
E_1\longleftrightarrow E_2\, .
\end{equation}
Thus, CY threefolds obtained as a hypersurface of $dP_2\times ...$ orientifolded by the above $\mathbb{Z}_2$ action will have $h^{1,1}_-=1$. $dP_{0,1}$ admit an analogous $\mathbb{Z}_2$ action, but the action on the divisor classes is trivial, so $h^{1,1}_-=0$.

$dP_3$ admits two distinct $\mathbb{Z}_2$ actions with co-dimension one fixed point locus. The first is the same as the one of $dP_2$, so the discussion is analogous. The $\mathbb{Z}_2$ fixed point locus is $\{x_{v_1}x_{e_2}-x_{v_2}x_{e_1}=0\}\cup \{x_{v_1}x_{e_2}+x_{v_2}x_{e_1}=x_3=0\}\cup \{x_{v_1}x_{e_2}+x_{v_2}x_{e_1}=x_{e_3}=0\}$, so it is the disjoint union of a $\mathbb{P}^1$ and two isolated fixed points. 

The second $\mathbb{Z}_2$ action is called $I_{dP_3}^{(2)}$ and acts as
\begin{equation}
x_{v_1}\longleftrightarrow x_{e_2}\, ,\quad x_{v_2}\longleftrightarrow x_{e_1}\, ,\quad x_{v_3}\longleftrightarrow x_{e_3}\, .
\end{equation}
The action on the divisor classes is 
\begin{align}
&l\longrightarrow 2l-\sum_{i=3}^3e_i\, ,\quad e_1\longrightarrow l-e_1-e_3\, ,\nonumber\\ &e_2\longrightarrow l-e_2-e_3\, ,\quad e_3\longrightarrow l-e_1-e_2\, .
\end{align}
This maps the K\"ahler form
\begin{equation}
J\longrightarrow J'=(2t^0-\sum_{i}t^i)l-(t^0-t^1-t^3)e_1-(t^0-t^2-t^3)e_2-(t^0-t^1-t^2)e_3\, ,
\end{equation}
which is in the K\"ahler cone if $J$ is in the K\"ahler cone. For $\mathbb{Z}_2$ invariance of the K\"ahler form we have to set $t^0=\sum_{i=1}^3 t^i$ and $t^1=t^2$. As the requirement of $\mathbb{Z}_2$ invariance fixes two linear combinations of the K\"ahler parameters, we have $h^{1,1}_-=2$.

Finally, $I_{dP_3}^{(3)}$ is defined by reflecting the fan over the origin, but the fixed point locus of this action is given by two constraints $x_1^2 x_{e_2}^2=x_2^2x_{e_3}^2=x_3^2x_{e_1}^2$, so it is of co-dimension two. It acts on the divisor classes as
\begin{align}
&l\longrightarrow 2l-\sum_{i=3}^3e_i\, ,\quad e_1\longrightarrow l-e_2-e_3\, ,\nonumber\\ &e_2\longrightarrow l-e_1-e_3\, ,\quad e_3\longrightarrow l-e_1-e_2\, .
\end{align}
This maps the K\"ahler form
\begin{equation}
J\longrightarrow J'=(2t^0-\sum_{i}t^i)l-(t^0-t^2-t^3)e_1-(t^0-t^1-t^3)e_2-(t^0-t^1-t^2)e_3\, ,
\end{equation}
which is also in the K\"ahler cone if $J$ is in the K\"ahler cone. Again, for $\mathbb{Z}_2$ invariance of the K\"ahler form we have to set $t^0=\sum_{i=1}^3 t^i$, so $h^{1,1}_-=1$.

\bibliographystyle{JHEP}
\bibliography{CicyRefs.bib}

\end{document}